\documentclass[twocolumn]{aastex702}

\usepackage{amsmath,amssymb,bm}
\usepackage{graphicx}
\usepackage{booktabs}
\usepackage{array}
\usepackage{placeins}

\newif\ifarxiv

\arxivtrue    

\ifarxiv
  \graphicspath{{./}} 
\else
  \graphicspath{{Figs/}}
\fi

\newcommand{\avg}[1]{\left\langle #1\right\rangle_\phi}
\newcommand{\kavg}{\avg{\kappa}}
\newcommand{\gpavg}{\avg{\gamma_+}}
\newcommand{\YMf}{Y_{\mathrm{MF}}}
\newcommand{\ksph}{\kappa_{\mathrm{sph}}}
\newcommand{\gsph}{\gamma_{+,\mathrm{sph}}}
\newcommand{\psimono}{\psi_{\mathrm{mono}}}
\newcommand{\epsp}{\epsilon_\perp}
\newcommand{\rtwo}{r_{200\mathrm c}}
\newcommand{\Mtwo}{M_{200\mathrm c}}
\newcommand{\deltag}{\delta_g}
\newcommand{\deltaY}{\delta_Y}
\newcommand{\deltamu}{\delta_\mu}
\newcommand{\deltamb}{\delta_{\mathrm{mb}}}
\newcommand{\Ckk}{C_{\kappa\kappa}}
\newcommand{\Ckplus}{C_{\kappa+}}
\newcommand{\Ckcross}{C_{\kappa\times}}
\newcommand{\Cplusplus}{C_{++}}
\newcommand{\Cpluscross}{C_{+\times}}
\newcommand{\Ccrosscross}{C_{\times\times}}
\newcommand{\Qoff}{Q_{\mathrm{off}}}
\newcommand{\phioff}{\varphi_{\mathrm{off}}}
\newcommand{\Nhalo}{N_\mathrm{halo}}

\begin{document}

\title{Why Azimuthal Averaging Works in Halo Lensing:\\ Symmetry and Power Counting for Nonlinear Shear and Magnification}

\author[0000-0002-7196-4822,gname=Keiichi,sname=Umetsu]{Keiichi Umetsu}
\affiliation{Academia Sinica Institute of Astronomy and Astrophysics (ASIAA), No. 1, Section 4, Roosevelt Road, Taipei 106319, Taiwan}
\email{keiichi@asiaa.sinica.edu.tw}

\shorttitle{Azimuthal Averaging in Halo Lensing}
\shortauthors{Umetsu}

\begin{abstract}
Cluster weak-lensing analyses often compress two-dimensional lensing fields into azimuthally averaged radial profiles and evaluate nonlinear observables from the averaged convergence and shear. However, the ring average of a nonlinear observable generally differs from the mean-field prediction formed from the averaged fields. This difference, divided by the mean-field prediction, defines the fractional residual. For a complete ring, the difference begins at second order in angular fluctuations. For centered elliptical halos, rotational symmetry makes the shape contribution even in signed ellipticity, while the leading miscentering dipole is orthogonal to the shape quadrupole. We test these predictions using projected triaxial Navarro--Frenk--White halos at six mass--redshift grid points spanning $3\leq M_{200\mathrm c}/(10^{14}\,h^{-1}M_\odot)\leq20$ and $0.2\leq z_l\leq0.5$, for $z_s=1$. For the reference offset model, centering offsets follow a Rayleigh distribution with scale $0.05\rtwo$. For $0.366\leq R/\rtwo\leq1$, where at least $95\%$ of each population satisfies a conservative subcriticality criterion, we calculate the median of the fractional residuals across the retained halos for each population and radius. The largest population-median magnitudes are $0.90\%$ for reduced shear, $0.0093\%$ for inverse magnification, $0.64\%$ for magnification, and $0.18\%$ across the two magnification-bias cases $\mu^{\alpha-1}$ ($\alpha=0.3, 1.4$). In the paired calculation at $z_s=2$, the maximum magnitude increases for every observable, consistent with weak-lensing power counting. Across both source planes and the common radial domain, every population median remains below $2\%$ in magnitude. This accuracy follows from first-order cancellation, rotational symmetry, and harmonic orthogonality, with further weak-lensing suppression of the remaining nonlinear terms.
\end{abstract}


\section{Introduction}
\label{sec:intro}

Gravitational lensing provides a direct probe of the projected matter distribution, with a geometric dependence on cosmological distances.  On cosmological scales, weak-lensing measurements constrain both the growth of structure and the expansion history of the Universe \citep[e.g.,][]{BartelmannSchneider2001}. Galaxy clusters occupy the deeply nonlinear regime of structure formation and serve as particularly powerful laboratories for astrophysics and cosmology, with weak and strong lensing offering complementary probes of their projected mass distributions \citep[e.g.,][]{KneibNatarajan2011,Umetsu2020rev}.

Weak gravitational lensing by galaxy clusters is often analyzed through one-dimensional radial profiles of the lensing signal measured from background galaxies.  The tangential shape distortion is represented by an azimuthally averaged reduced shear, while number-count depletion or enhancement is represented by an azimuthally averaged magnification-bias profile.  
This compression is statistically efficient, and the resulting radial lensing profiles directly constrain the projected mass distribution, providing a basis for cluster mass reconstruction and mass calibration \citep[e.g.,][]{Umetsu2020rev}.  

At the same time, galaxy clusters exhibit rich azimuthal structure in projection.  Triaxiality, centering offsets, substructure, correlated environment, and uncorrelated line-of-sight structure all generate angular variations in the convergence and shear.  Their effects on radial weak-lensing profiles and inferred cluster masses have been studied extensively \citep[e.g.,][]{Clowe2004,BeckerKravtsov2011,Grandis2021,Sommer2022}.  Azimuthal averaging itself does not require the projected cluster to be axisymmetric.

The potential problem arises from combining azimuthal averaging with nonlinear lensing observables.\footnote{We consider one source plane at a time and isolate azimuthal variations in the lensing fields.  Averaging over a source-redshift distribution is a distinct operation not considered here.}  We restrict our analysis to rings lying entirely in the subcritical region.  At a fixed source redshift, the reduced shear and inverse magnification at angular position $\bm{\theta}$ are
\begin{equation}
 \begin{aligned}
 g(\bm{\theta})&=\frac{\gamma(\bm{\theta})}{1-\kappa(\bm{\theta})},\\
 Y(\bm{\theta})&\equiv\mu^{-1}(\bm{\theta})
 =\left[1-\kappa(\bm{\theta})\right]^2-|\gamma(\bm{\theta})|^2.
 \end{aligned}
 \label{eq:main-basic-observables}
\end{equation}
Here the convergence $\kappa$ describes isotropic focusing, the complex shear $\gamma$ describes tidal distortion, and $\mu$ is the scalar magnification.  Magnification bias reflects two competing effects: the increase in observed solid angle reduces the source density, while flux amplification increases the number of sources detected above a fixed threshold.  
For the logarithmic slope $\alpha$ of the unlensed cumulative source counts, the response scales as $Y^{1-\alpha}$ \citep[e.g.,][]{Umetsu2020rev}.
For a non-axisymmetric lens there is no general identity requiring
\begin{equation}
 \avg{\mathcal{O}[\bm F(\phi)]}
 =
 \mathcal{O}[\avg{\bm F(\phi)}],
 \label{eq:main-noncommutativity}
\end{equation}
where $\bm F$ denotes the local lensing fields and the average is taken around a complete ring.  The observable averaged around the ring need not equal the mean-field prediction obtained by evaluating the same nonlinear map at the ring-averaged fields.

Two distinct effects must be distinguished in a one-dimensional analysis.
First, halo triaxiality or a displacement of the analysis center can change the azimuthally averaged lensing profiles themselves.  Second, the resulting angular variations can produce a difference between averaging the local nonlinear observable and evaluating it from the averaged fields.  Both terms in Equation~\eqref{eq:main-noncommutativity} are constructed from the same lensing fields about the same adopted center.  Consequently, changes in the azimuthally averaged lensing profiles are included in both terms, and the difference between them isolates the nonlinear-averaging effect studied here.

For reduced tangential shear, the corresponding quasi-circular approximation has been used previously \citep[e.g.,][]{Clowe2004,Umetsu2020rev}.  Using cluster halos from cosmological hydrodynamical simulations, \citet{Grandis2021} reported that forming reduced shear from the azimuthally averaged fields rather than averaging the local reduced shear changes the reduced-shear profile by only a few percent and shifts the inferred weak-lensing mass bias by $\lesssim1\%$.  The central question is thus why a difference that is not required to vanish remains small, and how its leading terms depend on halo shape, miscentering, and lensing strength.

We therefore address four questions.
\begin{enumerate}
\item Why can the ring average of a nonlinear halo-lensing observable be accurately approximated by the mean-field prediction formed from the ring-averaged lensing fields?
\item Which symmetries and power-counting arguments suppress the resulting difference?
\item How accurate is this approximation for reduced shear, inverse magnification, magnification, and magnification bias?
\item Which physical configurations limit this accuracy, and can those limits be predicted?
\end{enumerate}

To address these questions, we expand the nonlinear lensing observables about the mean field $\avg{\bm F}$ and combine the resulting moment hierarchy with symmetry arguments for projected halo shape and miscentering.  We test these predictions using populations of triaxial Navarro--Frenk--White (NFW) halos \citep{NavarroFrenkWhite1997} with specified distributions of centering offsets.

The paper is organized as follows.  In Section~\ref{sec:general-theory}, we develop the theory of nonlinear ring averaging.  In Section~\ref{sec:symmetry}, we derive additional suppression from rotational symmetry and the distinct angular structures of projected halo shape and miscentering.  In Section~\ref{sec:population}, we introduce the cluster-population experiment with projected triaxial halos and specified centering-offset distributions.  In Section~\ref{sec:accuracy}, we quantify the accuracy of the mean-field approximation for these observables and its dependence on lensing strength.  In Section~\ref{sec:limits}, we examine the limits and conditioning of the approximation, including the effects of correlations between centering-offset directions and projected halo shape.  In Section~\ref{sec:discussion}, we discuss the physical interpretation, scope, and limitations.  Finally, Section~\ref{sec:summary} summarizes our conclusions.

For the numerical calculations, we adopt the Planck 2018 spatially flat $\Lambda$ cold dark matter ($\Lambda$CDM) cosmology, with $\Omega_{\mathrm{m}}=0.3111$, $\Omega_{\mathrm{b}}=0.0490$, $\sigma_8=0.8102$, $n_s=0.9665$, and a Hubble constant $H_0=100h\,\mathrm{km\,s^{-1}\,Mpc^{-1}}$ with $h=0.6766$ \citep{PlanckCollaboration2020}.  The critical density at redshift $z$ is $\rho_{\mathrm{c}}(z)=3H^2(z)/(8\pi G)$, where $H(z)$ is the Hubble function.  We use $R$ for physical projected distance from the adopted analysis center and $r$ for three-dimensional halo-centric radius.  All radii are physical.  We define the spherical overdensity mass by $M_\Delta=(4\pi/3)\Delta\rho_{\mathrm{c}}(z)r_\Delta^3$, where the halo-centric radius $r_\Delta$ encloses a mean density $\Delta\rho_{\mathrm{c}}(z)$; $M_{200\mathrm c}$ and $r_{200\mathrm c}$ correspond to $\Delta=200$.

\section{General theory of nonlinear ring averaging}
\label{sec:general-theory}

\subsection{Lensing fields, ideal ring average, and the target residual}
\label{subsec:main-setup}

Let $\bm{\theta}$ denote angular position on the sky relative to the adopted analysis center, with angular radius $\theta\equiv|\bm{\theta}|$ and polar angle $\phi$.  For a lens at redshift $z_l$, the corresponding physical projected radius is $R=D_l\theta$, where $D_l$ is the observer--lens angular diameter distance.  For a source plane at redshift $z_s>z_l$, the convergence and complex shear are
\begin{equation}
 \begin{aligned}
 \kappa(R,\phi) &\equiv\frac{\Sigma(R,\phi)}{\Sigma_{\mathrm{cr}}},\\
 \gamma(R,\phi) &\equiv\gamma_1(R,\phi)+i\gamma_2(R,\phi),
  \end{aligned}
 \label{eq:main-kappa-gamma-definitions}
\end{equation}
where
\begin{equation}
 \Sigma_{\mathrm{cr}}
 =\frac{c^2D_s}{4\pi G D_lD_{ls}},
 \label{eq:main-sigma-critical}
\end{equation}
and $D_s$ and $D_{ls}$ are the observer--source and lens--source angular diameter distances, respectively.  Let $\bm{\theta}_s$ denote the corresponding angular position in the source plane.  For lensing generated by a scalar potential, the Jacobian of the local lens mapping,
\begin{equation}
 \mathcal{A}
 \equiv
 \frac{\partial\bm{\theta}_s}{\partial\bm{\theta}}
 =
 \begin{pmatrix}
  1-\kappa-\gamma_1 & -\gamma_2\\
  -\gamma_2 & 1-\kappa+\gamma_1
 \end{pmatrix},
 \label{eq:main-lensing-jacobian}
\end{equation}
is symmetric.  Its determinant and eigenvalues are
\begin{equation}
 \begin{aligned}
 \det\mathcal{A}
 &=(1-\kappa)^2-|\gamma|^2
 =\mu^{-1}=Y,\\
 \lambda_\pm &=1-\kappa\pm|\gamma|,
 \end{aligned}
 \label{eq:main-jacobian-properties}
\end{equation}
where $Y$ is the inverse magnification introduced in Equation~\eqref{eq:main-basic-observables}.  
We restrict our analysis of ring-averaged observables to complete rings lying in the subcritical region, with $\lambda_-(R,\phi)>0$ at every azimuth.  This condition ensures $1-\kappa>0$, $Y>0$, and $|g|<1$ throughout each ring.  The numerical calculations impose the stronger far-background subcriticality criterion in Equation~\eqref{eq:main-safety}.

In the local polar basis, the tangential and cross shear components are defined by
\begin{equation}
 \gamma_+
 =-\Re\!\left(\gamma e^{-2i\phi}\right),
 \qquad
 \gamma_\times
 =-\Im\!\left(\gamma e^{-2i\phi}\right).
 \label{eq:main-radial-shear-definitions}
\end{equation}
The symmetric Jacobian is thus fully specified in the radial basis by the three local fields $\bm F=(\kappa,\gamma_+,\gamma_\times)$, from which the nonlinear observables considered below are constructed.

At fixed projected radius $R$, $\avg{\bm F(R,\phi)}$ denotes the fields averaged over a complete, uniformly weighted ring about the adopted analysis center.  Throughout this paper, $\avg{\cdot}$ denotes this azimuthal operation at fixed $R$ and $z_s$.
The mean-field prediction for an observable $\mathcal{O}$ is
\begin{equation}
 \mathcal{O}_{\mathrm{MF}}(R)
 \equiv
 \mathcal{O}\!\left[\avg{\bm F(R,\phi)}\right].
 \label{eq:main-observable-definitions}
\end{equation}
The exact ring average is $\avg{\mathcal{O}[\bm F(R,\phi)]}$, and the residual studied here is
\begin{equation}
 \Delta_\phi\mathcal{O}(R)
 \equiv
 \avg{\mathcal{O}[\bm F(R,\phi)]}
 -\mathcal{O}_{\mathrm{MF}}(R).
 \label{eq:main-general-residual}
\end{equation}
Both terms are evaluated about the same adopted center and at the same source redshift.  Finite radial bins, angular masks, nonuniform source weights, and discrete source sampling are not included in this ideal-ring calculation; their effects are discussed in Section~\ref{subsec:discussion-scope}.

\subsection{Moment hierarchy, first-order cancellation, and power counting}
\label{subsec:main-first-order}

We decompose the local lensing fields $\bm F=(\kappa,\gamma_+,\gamma_\times)$ into their ring means and angular fluctuations:
\begin{equation}
 \bm F(R,\phi)
 =\avg{\bm F(R,\phi)}+\delta\bm F(R,\phi),
 \qquad
 \avg{\delta\bm F}=0.
 \label{eq:main-mean-fluctuation}
\end{equation}
Here $a\in\{\kappa,+,\times\}$ labels the field components, and $\delta F_a$ is the angular fluctuation of $F_a$ about its ring mean. The corresponding central-moment tensors are
\begin{equation}
 M_{a_1\cdots a_n}(R)
 \equiv
 \avg{\delta F_{a_1}\cdots\delta F_{a_n}}.
 \label{eq:main-central-moments}
\end{equation}
For $n=2$, $C_{ab}\equiv M_{ab}=\avg{\delta F_a\delta F_b}$ defines the angular covariance entries.  Throughout, $M_{a_1\cdots a_n}$ and $C_{ab}$ denote azimuthal moments within a single ring at fixed $R$ and $z_s$; the halo-population statistics are introduced separately in Section~\ref{sec:accuracy}.  Here and below, $\mathrm{Var}_\phi$ denotes the azimuthal variance about the ring mean at fixed $R$ and $z_s$; for example, $\mathrm{Var}_\phi(Y)\equiv\avg{\bigl(Y-\avg{Y}\bigr)^2}$.

Whenever the Taylor expansion of a local observable about $\avg{\bm F}$ converges over the angular excursions on the ring, the residual defined in Equation~\eqref{eq:main-general-residual} has the moment expansion
\begin{equation}
 \begin{aligned}
 \Delta_\phi\mathcal{O}
 &=
 \sum_{n=2}^{\infty}\frac{1}{n!}
 \sum_{a_1,\ldots,a_n}
 \frac{\partial^n\mathcal{O}}
 {\partial F_{a_1}\cdots\partial F_{a_n}}
 \!\left(\avg{\bm F}\right)
 M_{a_1\cdots a_n}\\
 &=
 \frac12\sum_{a,b}
 \frac{\partial^2\mathcal{O}}
 {\partial F_a\,\partial F_b}
 \!\left(\avg{\bm F}\right)
 C_{ab}
 +O(\delta F^3).
 \end{aligned}
 \label{eq:main-residual-hierarchy}
\end{equation}
The $n=1$ term is absent because $\avg{\delta\bm F}=0$. The ring average is therefore organized as the mean-field term plus a correction that begins at second order in the angular fluctuations, independent of any particular halo model or weak-lensing approximation. At leading order, this correction is a Hessian--covariance contraction. Its magnitude and sign depend jointly on the angular covariance and the local curvature of $\mathcal{O}$ at $\avg{\bm F}$; an indefinite Hessian permits cancellations between directions of opposite curvature. Appendix~\ref{app:general-observables} collects the complete set of second-order angular covariance entries, together with several useful bounds and special cases.

When $\mathcal{O}_{\mathrm{MF}}\neq0$, we define the corresponding fractional residual by
\begin{equation}
 \delta_{\mathcal{O}}(R)
 \equiv
 \frac{\Delta_\phi\mathcal{O}(R)}{\mathcal{O}_{\mathrm{MF}}(R)}
 =
 \frac{\avg{\mathcal{O}[\bm F(R,\phi)]}}
      {\mathcal{O}_{\mathrm{MF}}(R)}
 -1,
 \label{eq:main-general-fractional-residual}
\end{equation}
where $\delta_{\mathcal{O}}$ denotes the dimensionless residual after ring averaging. Its denominator is the mean-field prediction, not the exact ring average. Thus $\delta_{\mathcal{O}}$ measures the fractional residual relative to the mean-field baseline $\mathcal{O}_\mathrm{MF}$ and should not be interpreted as a conventional fractional bias normalized to the ring average $\avg{\mathcal{O}[\bm F]}$.

The residual vanishes identically in two special cases. For an axisymmetric field, $\delta\bm F=0$, so that the equality holds at arbitrary lensing strength. For a linear observable, ring averaging commutes with the observable even for a non-axisymmetric field.

For power counting, let $\eta$ characterize the overall amplitude of the lensing fields and $\epsilon$ their characteristic fractional anisotropy, such that
\begin{equation}
 \kavg,\gpavg=O(\eta),
 \qquad
 \delta\kappa,\delta\gamma_+,\delta\gamma_\times
 =O(\eta\epsilon).
 \label{eq:main-eta-epsilon}
\end{equation}
Here $\epsilon$ characterizes the angular structure without specifying its physical origin.  The corresponding physical perturbation scales for projected shape and miscentering are identified in Section~\ref{sec:symmetry}.

It follows that $C_{ab}=O(\eta^2\epsilon^2)$. If the derivatives of $\mathcal{O}$ remain regular, the residual $\Delta_\phi\mathcal{O}$ is generically $O(\eta^2\epsilon^2)$. The scaling of the fractional residual also depends on the leading amplitude of $\mathcal{O}_{\mathrm{MF}}$. For an observable with $\mathcal{O}_{\mathrm{MF}}=O(\eta)$, such as reduced shear,
\begin{equation}
 \delta_{\mathcal{O}}=O(\eta\epsilon^2).
 \label{eq:main-power-counting}
\end{equation}
An anomalously small $\mathcal{O}_{\mathrm{MF}}$ can instead amplify the fractional residual, as discussed in Section~\ref{subsec:main-conditioning}. The weak-lensing regime therefore provides an additional suppression beyond the cancellation of the first angular moment.

\subsection{Reduced tangential shear}
\label{subsec:main-reduced-shear}

For reduced tangential shear, the mean-field prediction is 
\begin{equation}
 g_{+,\mathrm{MF}}
 \equiv
 \frac{\gpavg}{1-\kavg}
 =
 \frac{\gpavg}{D},
 \qquad
 D\equiv1-\kavg.
 \label{eq:main-gmf-definition}
\end{equation}
For any complete circular ring about the adopted analysis center, the azimuthally averaged tangential shear obeys the monopole relation \citep{Kaiser1995}:
\begin{equation}
 \gpavg
 =\overline{\kappa}(<R)-\kavg,
 \label{eq:main-tangential-shear-monopole}
\end{equation}
where
\begin{equation}
 \overline{\kappa}(<R)
 \equiv
 \frac{2}{R^2}\int_0^R dR'\,R'\avg{\kappa(R',\phi)}
 \label{eq:main-interior-mean-convergence}
\end{equation}
is the mean convergence interior to $R$ about the same adopted center.  An overline denotes an interior radial average or moment, whereas $\avg{\cdot}$ denotes an azimuthal ring mean.
Equation~\eqref{eq:main-tangential-shear-monopole} is an identity for a scalar lensing potential and does not require axisymmetry about the adopted center.

Using the general definition in Equation~\eqref{eq:main-general-fractional-residual}, we define the fractional residual for $g_+$ by
\begin{equation}
 \deltag(R)
 \equiv
 \frac{\avg{g_+}}{g_{+,\mathrm{MF}}}-1.
 \label{eq:main-deltag-definition}
\end{equation}
For explicit field labels $A,B\in\{\kappa,+,\times\}$, we write the angular covariance entries as
\begin{equation}
 C_{AB}(R)
 \equiv
 \avg{\delta F_A\,\delta F_B}.
\label{eq:main-CAB-definition}
\end{equation}
Applying the general moment expansion in Equation~\eqref{eq:main-residual-hierarchy} to $g_+=\gamma_+/(1-\kappa)$ gives
\begin{equation}
 \Delta_\phi g_+
 \equiv
 \avg{g_+}-g_{+,\mathrm{MF}}
 =
 \frac{\Ckplus}{D^2}
 +\frac{\gpavg\Ckk}{D^3}
 +O(\delta F^3).
 \label{eq:main-delta-g}
\end{equation}
For $\gpavg\neq0$, dividing by $g_{+,\mathrm{MF}}$ gives the second-order fractional residual
\begin{equation}
 \deltag^{(2)}(R)
 =
 \frac{\Ckplus}{D\gpavg}
 +\frac{\Ckk}{D^2}.
 \label{eq:main-deltag-second-order}
\end{equation}
Equation~\eqref{eq:main-deltag-second-order} gives the leading correction to the quasi-circular approximation \citep{Clowe2004,Umetsu2020rev}.  The term involving the convergence--tangential-shear covariance $\Ckplus$ has no universal sign, so that the residual is not simply a positive variance correction.

Equations~\eqref{eq:main-delta-g} and \eqref{eq:main-deltag-second-order}, together with Equation~\eqref{eq:main-power-counting}, also distinguish a large value of $\Delta_\phi g_+$ from a large fractional residual $\deltag$ that arises when $g_{+,\mathrm{MF}}$ is close to zero.  This distinction becomes important in Section~\ref{subsec:main-conditioning}.  A model-independent bound on $\Delta_\phi g_+$ is given in Appendix~\ref{app:general-observables}.

\subsection{Cross shear and reduced cross shear}
\label{subsec:main-cross}

Using the angular radius $\theta$ and the standard dimensionless lensing potential $\psi(\theta,\phi)$, the cross component in the local radial frame is
\begin{equation}
 \gamma_\times(\theta,\phi)
 =-\frac{\partial}{\partial\theta}
 \left(\frac{1}{\theta}\frac{\partial\psi}{\partial\phi}\right).
 \label{eq:main-cross-total-derivative}
\end{equation}
Its complete-ring mean is thus
\begin{equation}
 \avg{\gamma_\times(\theta,\phi)}
 =-\frac{\partial}{\partial\theta}
 \left[\frac{1}{2\pi\theta}
 \int_0^{2\pi}d\phi\,\frac{\partial\psi}{\partial\phi}\right]
 =0.
 \label{eq:main-true-cross-zero}
\end{equation}
The result holds about an arbitrary adopted origin; axisymmetry is not required \citep{Kaiser1995}.  Galaxy shapes respond to the reduced cross shear
\begin{equation}
 g_\times=\frac{\gamma_\times}{1-\kappa}.
\end{equation}
Its azimuthal average is
\begin{equation}
 \avg{g_\times}
 =
 \frac{\Ckcross}{D^2}
 +O(\delta F^3),
 \qquad
 \Ckcross=\avg{\delta\kappa\,\delta\gamma_\times}.
 \label{eq:main-reduced-cross}
\end{equation}
The mean-field prediction is $g_{\times,\mathrm{MF}}=\avg{\gamma_\times}/D=0$, whereas $\avg{g_\times}$ need not vanish because the azimuthally varying factor $1/(1-\kappa)$ can correlate with $\gamma_\times$.  For a centered halo with reflection symmetry about its projected principal axes, $g_\times$ is odd under reflection and hence $\avg{g_\times}=0$ exactly; in particular, $\Ckcross=0$.  The ensemble expectation of $\avg{g_\times}$ also vanishes in a population for which every configuration and its mirror image have equal statistical weight.

A generic translated elliptical halo need not possess a reflection axis about the adopted center, so that $\avg{g_\times}$ need not vanish for an individual halo. Appendix~\ref{app:general-observables} gives the complete set of second-order angular covariance entries involving $\gamma_\times$ and the leading variance of the reduced cross shear.

\subsection{Inverse magnification, magnification, and magnification bias}
\label{subsec:main-magnification}

The inverse magnification $Y=\mu^{-1}$ is quadratic in the local lensing fields:
\begin{equation}
 Y
 =(1-\kappa)^2-\gamma_+^2-\gamma_\times^2,
 \qquad
 \YMf\equiv D^2-\gpavg^2.
 \label{eq:main-Y-definitions}
\end{equation}
Using Equation~\eqref{eq:main-CAB-definition}, the angular shear variances entering inverse magnification are $\Cplusplus=\avg{\delta\gamma_+^2}$ and $\Ccrosscross=\avg{\gamma_\times^2}$. Using the mean--fluctuation decomposition in Equation~\eqref{eq:main-mean-fluctuation}, the ring average of $Y$ is
\begin{equation}
 \begin{aligned}
 \avg{Y}&=\YMf+\Delta_\phi Y,\\
 \Delta_\phi Y&=\Ckk-\Cplusplus-\Ccrosscross.
 \end{aligned}
 \label{eq:main-Y-exact}
\end{equation}
Magnification and magnification bias apply a further nonlinear transformation to $Y$.  Magnification is $\mu=Y^{-1}$.  For magnification bias at an adopted flux cut $F_{\mathrm{cut}}$, define $\alpha\equiv-\left.d\ln n_{\mathrm{unl}}(>F)/d\ln F\right|_{F=F_{\mathrm{cut}}}$, where $n_{\mathrm{unl}}(>F)$ is the unlensed cumulative source surface density above flux $F$.\footnote{The corresponding magnitude-count slope is conventionally denoted by $s\equiv d\log_{10}n_{\mathrm{unl}}(<m)/dm$, evaluated at the adopted magnitude cut $m_{\mathrm{cut}}$ corresponding to $F_{\mathrm{cut}}$, so that $\alpha=2.5s$.  In the faint-end power-law limit of a Schechter luminosity function, $\alpha=-(\alpha_{\mathrm{LF}}+1)$, where $\alpha_{\mathrm{LF}}$ is the faint-end luminosity-function slope.}  
Approximating the counts by a power law over the relevant flux range, we set $q\equiv1-\alpha$. The local number-count response is thus
\begin{equation}
 n_\mu\propto\mu^{\alpha-1}=Y^q.
 \label{eq:main-magnification-bias-map}
\end{equation}
The subcritical restriction ensures $Y(R,\phi)>0$ and $\YMf(R)>0$, so that the real powers and logarithms used below are well defined.

The mean-field baselines for inverse magnification, magnification, and magnification bias are $\YMf$, $\mu_{\mathrm{MF}}\equiv\YMf^{-1}$, and $\YMf^q$, respectively.  Using the general definition in Equation~\eqref{eq:main-general-fractional-residual}, we define the corresponding fractional residuals by
\begin{equation}
 \begin{aligned}
 \deltaY
 &\equiv
 \frac{\avg{Y}}{\YMf}-1,\\
 \deltamu
 &\equiv
 \frac{\avg{\mu}}{\mu_{\mathrm{MF}}}-1
 =\frac{\avg{Y^{-1}}}{\YMf^{-1}}-1,\\
 \deltamb(q)
 &\equiv
 \frac{\avg{Y^q}}{\YMf^q}-1,
 \qquad q=1-\alpha.
 \end{aligned}
 \label{eq:main-magnification-residuals}
\end{equation}
Thus, the $Y^q$ family contains inverse magnification at $q=1$, magnification at $q=-1$, and magnification bias at $q=1-\alpha$.  Combining Equations~\eqref{eq:main-Y-exact} and \eqref{eq:main-magnification-residuals} gives the inverse-magnification residual
\begin{equation}
 \deltaY
 =\frac{\Delta_\phi Y}{\YMf}
 =\frac{\Ckk-\Cplusplus-\Ccrosscross}{\YMf}.
 \label{eq:main-deltaY-exact}
\end{equation}
The absence of higher-order terms does not by itself guarantee that $\deltaY$ is small: $\Delta_\phi Y$ has no fixed sign, and the ratio can be amplified where $\YMf$ is small.

For general $q$, the comparison with the mean-field prediction factorizes as
\begin{equation}
 \frac{\avg{Y^q}}{\YMf^q}
 =
 (1+\deltaY)^q
 \frac{\avg{Y^q}}{\bigl(\avg{Y}\bigr)^q}.
 \label{eq:main-Yq-factorization}
\end{equation}
The factorization can be summarized by $\YMf^q\longrightarrow\bigl(\avg{Y}\bigr)^q\longrightarrow\avg{Y^q}$.  The first factor describes how the difference between the ring average $\avg{Y}$ and its mean-field prediction $\YMf$ affects the power-law response.  The second isolates the additional effect of applying the nonlinear map $Y\mapsto Y^q$ before, rather than after, azimuthal averaging.

To describe the second factor at all orders, let $Z(\phi)\equiv\ln Y(\phi)$ and define the angular cumulant-generating function
\begin{equation}
 K_Z(t)\equiv\ln\avg{e^{tZ}}=\ln\avg{Y^t}.
\end{equation}
Taking the logarithm of the second factor in Equation~\eqref{eq:main-Yq-factorization} gives
\begin{equation}
 \ln\frac{\avg{Y^q}}{\bigl(\avg{Y}\bigr)^q}
 =K_Z(q)-qK_Z(1).
 \label{eq:main-CGF}
\end{equation}
Here the first-cumulant terms in $K_Z(q)$ and $qK_Z(1)$ are both $q\avg{Z}$ and they cancel identically.  The logarithm of the second factor thus depends only on the angular variance and higher cumulants of $Z=\ln Y$. For $Y>0$, Jensen's inequality gives $\avg{Y^q}\leq\bigl(\avg{Y}\bigr)^q$ for $0<q<1$ and $\avg{Y^q}\geq\bigl(\avg{Y}\bigr)^q$ for $q<0$ or $q>1$. The logarithm of the total ratio $\avg{Y^q}/\YMf^q$ also contains the first-factor contribution $q\ln(1+\deltaY)$. Since $\deltaY$ can have either sign, this contribution can reinforce or oppose the second-factor contribution. The total fractional residual $\deltamb(q)$ thus has no fixed sign.

To second order, the fractional residual for a $Y^q$ observable is
\begin{equation}
 \deltamb(q)
 =
 q\frac{\Delta_\phi Y}{\YMf}
 +\frac{q(q-1)}{2}
  \frac{\mathrm{Var}_\phi(Y)}{\YMf^2}
 +O(\delta F^3).
 \label{eq:main-Yq-second-order}
\end{equation}
For $q=1$, the variance term and all higher-order terms vanish, so that Equation~\eqref{eq:main-Yq-second-order} reduces to the inverse-magnification residual in Equation~\eqref{eq:main-deltaY-exact}.  
Setting $q=-1$ gives the second-order magnification residual $\deltamu$.

These expressions concern azimuthal averaging around an individual lens at fixed radius and source redshift.  Reduced-shear and magnification-bias corrections have also been studied for ensemble statistics of the cosmological shear field \citep[e.g.,][]{Shapiro2009,KrauseHirata2010,Deshpande2020}.  For nonlinear cluster lensing, source-redshift averaging was formulated for reduced shear by \citet{SeitzSchneider1997} and for inverse magnification and magnification bias by \citet{Umetsu2013}; see \citet{Umetsu2020rev} for a review of both cases.

Averaging over a source-redshift distribution is a separate operation and is not treated here.  Appendix~\ref{app:general-observables} develops the cumulant expansion of Equation~\eqref{eq:main-CGF}, the variance contribution to the second factor, and the Gaussian and lognormal special cases.

\section{Symmetry constraints from projected halo shape and miscentering}
\label{sec:symmetry}

The generic first-order cancellation derived above does not depend on the origin of the angular fluctuations $\delta\bm F$.  Projected halo shape and miscentering obey additional rotational and harmonic symmetries that further suppress the residual.  In this section, we identify their leading angular modes and use these symmetries to derive the leading dependence of the residual on projected ellipticity, centering offset, and relative orientation.

Appendix~\ref{app:coherent-halo} develops the projected-shape calculation from potential multipoles to the area-preserving elliptical profile, while Appendix~\ref{app:miscentering-alignment} gives the detailed derivations for miscentering and shape--centering coupling.

\subsection{Projected shape of a centered halo: an \texorpdfstring{$m=2$}{m=2} perturbation}
\label{subsec:main-shape}

Let $q_\perp$ denote the projected minor-to-major axis ratio and define the projected ellipticity $\epsp\equiv(1-q_\perp)/(1+q_\perp)$.\footnote{The commonly used alternative ellipticity $e\equiv(1-q_\perp^2)/(1+q_\perp^2)$ satisfies $e/2=\epsp+O(\epsp^3)$; thus, at linear order, $e=2\epsp$.}  The leading response to projected halo ellipticity is linear in $\epsp$ and quadrupolar, as commonly used in weak-lensing halo-shape analyses \citep[for review, see][]{Umetsu2020rev}.

We measure the polar angle $\phi$ from the projected major axis, so that $\phi=0$ lies along that axis.  For a centered elliptical halo, reflection symmetry about the projected principal axes makes $\delta\kappa$ and $\delta\gamma_+$ even in $\phi$, while $\gamma_\times$ is odd.  We define the radial amplitudes of the $m=2$ Fourier modes by
\begin{equation}
 \begin{aligned}
 F_2^\kappa(R)&\equiv2\avg{\delta\kappa(R,\phi)\cos2\phi},\\
 F_2^+(R)&\equiv2\avg{\delta\gamma_+(R,\phi)\cos2\phi},\\
 F_2^\times(R)&\equiv2\avg{\gamma_\times(R,\phi)\sin2\phi}.
 \end{aligned}
 \label{eq:main-m2-amplitudes}
\end{equation}
These definitions extract the $m=2$ coefficients from the local lensing fields $\bm F(R,\phi)$.  At leading order in $\epsp$, the fields are described by these quadrupole terms:
\begin{equation}
 \begin{aligned}
 \delta\kappa(R,\phi)&\simeq F_2^\kappa(R)\cos2\phi,\\
 \delta\gamma_+(R,\phi)&\simeq F_2^+(R)\cos2\phi,\\
 \gamma_\times(R,\phi)&\simeq F_2^\times(R)\sin2\phi.
 \end{aligned}
 \label{eq:main-m2-fields}
\end{equation}
All three amplitudes vanish in the circular limit and are linear in $\epsp$ at leading order.  In the power counting of Equation~\eqref{eq:main-eta-epsilon}, the resulting shape-induced field fluctuations are $O(\eta\epsp)$, so that $\epsilon=O(\epsp)$.

Let $\ksph(R)$ denote the convergence profile of the centered spherical reference halo.  For a small area-preserving elliptical deformation of this profile, the leading convergence quadrupole amplitude is
\begin{equation}
 F_2^\kappa(R)
 =-\epsp\frac{d\ksph(R)}{d\ln R}
 +O(\epsp^3).
 \label{eq:main-K2-slope}
\end{equation}
Thus, $F_2^\kappa$ is determined locally by the projected ellipticity and the radial slope of the convergence profile.  The shear amplitudes $F_2^+$ and $F_2^\times$ also depend on the enclosed mass distribution.  The three amplitudes are linked because they arise from the same quadrupole of the lensing potential: Appendix~\ref{appB:potential-closure} derives this relation, and Appendix~\ref{appB:elliptical-profile} gives their leading expressions for the area-preserving elliptical profile.

\subsection{Axis interchange and even-power suppression}
\label{subsec:main-evenness}

Beyond the generic first-order cancellation in Equation~\eqref{eq:main-residual-hierarchy}, axis interchange imposes an additional constraint on scalar residuals for centered elliptical halos.  To expose this symmetry, we formally allow $\epsp$ to take either sign,
\begin{equation}
 q_\perp(\epsp)=\frac{1-\epsp}{1+\epsp}.
\end{equation}
Under $\epsp\to-\epsp$, the axis ratio transforms as $q_\perp\to q_\perp^{-1}$.  This transformation interchanges the projected major and minor axes and maps the halo to the same projected configuration rotated by $\pi/2$.

Since a complete ring is unchanged by rotation, both the ring average of any scalar lensing observable $\mathcal{S}$ and its corresponding mean-field prediction are invariant under $\epsp\to-\epsp$:
\begin{equation}
\begin{aligned}
 \avg{\mathcal{S}}(\epsp)
 &=\avg{\mathcal{S}}(-\epsp),\\
 \mathcal{S}_{\mathrm{MF}}(\epsp)
 &=\mathcal{S}_{\mathrm{MF}}(-\epsp).
\end{aligned}
\label{eq:main-ring-evenness}
\end{equation}
The corresponding fractional residual $\delta_{\mathcal{S}}$, defined by Equation~\eqref{eq:main-general-fractional-residual} with $\mathcal{O}=\mathcal{S}$, is therefore even in $\epsp$.  In the circular limit, the centered halo has no angular structure, so that $\avg{\mathcal{S}}(0)=\mathcal{S}_{\mathrm{MF}}(0)$ and $\delta_{\mathcal{S}}(0)=0$.  If $\delta_{\mathcal{S}}$ is analytic about $\epsp=0$, its expansion consequently contains only even powers beginning at second order:
\begin{equation}
 \delta_{\mathcal{S}}(\epsp)
 =A_2\epsp^2+A_4\epsp^4+\cdots.
 \label{eq:main-even-ellipticity}
\end{equation}
This is an all-orders symmetry statement: the expansion about the circular limit excludes every odd power of $\epsp$, not only the linear term.  Appendix~\ref{appB:elliptical-profile} demonstrates the symmetry explicitly for the area-preserving elliptical profile used in our calculations.

\subsection{Miscentering: an \texorpdfstring{$m=1$}{m=1} translation}
\label{subsec:main-miscentering}

Offsets between the adopted center and the true halo center are commonly included in radial cluster-lensing models \citep[e.g.,][]{Grandis2021,Sommer2022}.  Unlike projected ellipticity $\epsp$, whose signed extension represents interchange of the projected axes, miscentering is described by a two-dimensional translation vector.

For a spherical halo, let $\bm{d}$ point from the adopted center to the true halo center and define its magnitude by $d\equiv|\bm{d}|\geq0$.  Choosing $\phi=0$ along $\bm{d}$, the true halo-centric projected radius $R_{\mathrm{h}}$ of a point with adopted-center polar coordinates $(R,\phi)$ satisfies
\begin{equation}
R_{\mathrm{h}}^2=R^2+d^2-2Rd\cos\phi.
\label{eq:main-displaced-radius}
\end{equation}
Miscentering changes the origin about which the same halo is described and averaged.  Consequently, even a spherical halo acquires angular structure when expressed about the adopted center.

We place realization-dependent parameters after a vertical bar and leave the remaining halo parameters implicit; thus $\bm F(R,\phi\mid d)$ denotes the local lensing fields evaluated about the adopted center for an offset of magnitude $d$. Let $\ksph(R)$ and $\gsph(R)$ denote the convergence and tangential shear of the same centered spherical reference halo. With primes denoting derivatives with respect to $R$, the small-offset expansion for $d/R\ll1$ gives
\begin{equation}
\begin{aligned}
\delta\kappa(R,\phi\mid d)&=-d\,\ksph'(R)\cos\phi+O(d^2),\\
\delta\gamma_+(R,\phi\mid d)&=-d\,\gsph'(R)\cos\phi+O(d^2),\\
\gamma_\times(R,\phi\mid d)&=2\frac{d}{R}\gsph(R)\sin\phi+O(d^2).
\end{aligned}
\label{eq:main-m1-fields}
\end{equation}
The first two expressions arise from translating the radial profiles, while the cross component arises because the tangential direction defined about the true halo center differs from that defined about the adopted center. Their $\cos\phi$ and $\sin\phi$ dependence shows that the leading translation-induced perturbation has angular mode $m=1$.

The natural local expansion variable is therefore
\begin{equation}
u\equiv\frac{d}{R}\geq0.
\label{eq:main-u-definition}
\end{equation}
The leading offset-induced field fluctuations in Equation~\eqref{eq:main-m1-fields} are $O(\eta u)$, so that $\epsilon=O(u)$ in the power counting of Equation~\eqref{eq:main-eta-epsilon}.  For a spherical halo, rotating the offset vector $\bm{d}$ leaves both the ring average $\avg{\mathcal{S}}$ and its mean-field prediction $\mathcal{S}_{\mathrm{MF}}$ unchanged.  If the resulting residual $\delta_{\mathcal{S}}$ is analytic in the components of $\bm{d}$ at $\bm{d}=0$, rotational invariance permits only even powers of $d$.  This is the translation counterpart of the even-power suppression under axis interchange derived in Equation~\eqref{eq:main-even-ellipticity}.  Thus, the first-order $m=1$ perturbations average to zero over a complete ring, and the first allowed offset contribution to $\delta_{\mathcal{S}}$ is $O(d^2)$, or equivalently $O(u^2)$ at fixed $R$, as derived in Section~\ref{subsec:main-interaction}.

Even when $d/\rtwo\ll1$, the offset need not be a small perturbation on rings with $R\sim d$, where the local expansion parameter $u=d/R$ is of order unity.

Projected shape and miscentering generate different leading angular modes:
\begin{equation}
\text{miscentering}: m=1,
\qquad
\text{projected shape}: m=2.
\label{eq:main-m1-m2}
\end{equation}
This harmonic distinction underlies the leading shape--centering orthogonality discussed next.

\begin{figure}[t]
\centering
\includegraphics[width=\columnwidth]{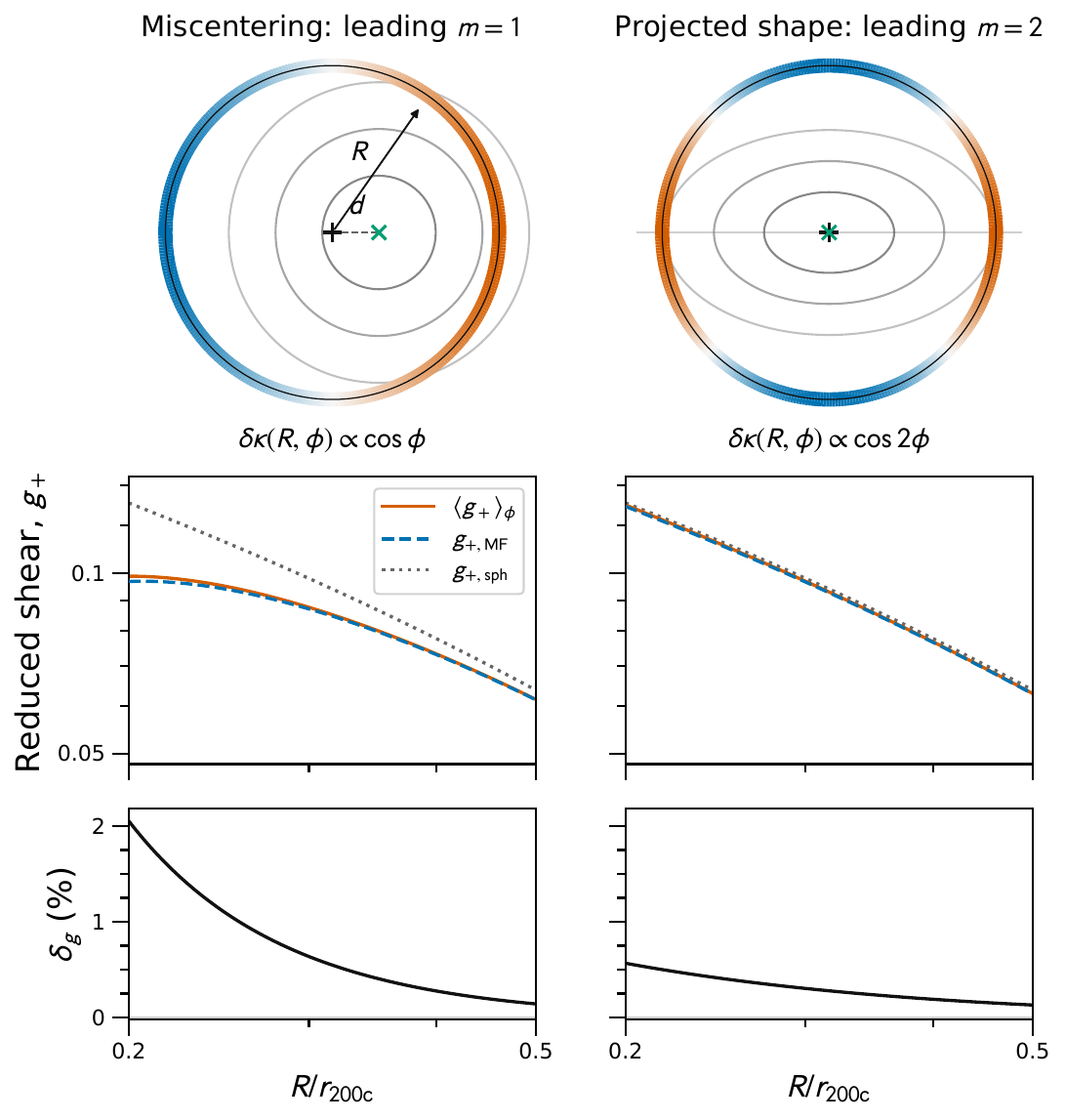}
\caption{Leading angular structures and nonlinear ring averaging for miscentering (left; $m=1$) and projected halo shape (right; $m=2$). All panels use $M_{200\mathrm c}=10^{15}h^{-1}M_\odot$, NFW concentration $c_{200\mathrm c}=3.84$, $z_l=0.3$, and $z_s=1$. The left column shows a spherical halo offset by $d=0.10\rtwo$ from the adopted center; the black plus and green cross mark the adopted and true centers, respectively. The right column shows a centered elliptical halo with $q_\perp=0.67$ ($\epsp\simeq0.20$). The top panels schematically show the corresponding $\kappa$ fluctuations. The middle panels compare $\avg{g_+}$, obtained by averaging $g_+(R,\phi)$, with $g_{+,\mathrm{MF}}$, obtained from $\avg{\kappa}$ and $\avg{\gamma_+}$; $g_{+,\mathrm{sph}}$ is the centered spherical-equivalent reference. The bottom panels show $\deltag=\avg{g_+}/g_{+,\mathrm{MF}}-1$ in percent. Separation from the spherical reference reflects radial-profile changes, whereas the difference between $\avg{g_+}$ and $g_{+,\mathrm{MF}}$ isolates nonlinear averaging.}
\label{fig:main-ring-averaging}
\end{figure}

Figure~\ref{fig:main-ring-averaging} illustrates these isolated $m=1$ and $m=2$ configurations and their reduced-shear residuals $\deltag$.


\subsection{Harmonic orthogonality and the reduced-shear shape--centering interaction}
\label{subsec:main-interaction}

Hydrodynamical simulations show that centering offsets need not have random directions relative to projected cluster structure \citep{Sommer2024,Sommer2025}.  Motivated by this result, we retain the orientation variable $\phioff$, defined as the angle between the offset direction and the projected major axis.  We formulate the shape--centering interaction for the reduced-shear residual $\deltag$ and give the corresponding parity-odd result for the ring-averaged reduced cross shear $\avg{g_\times}$.

\begin{figure*}[t]
\centering
\includegraphics[width=0.9\textwidth]{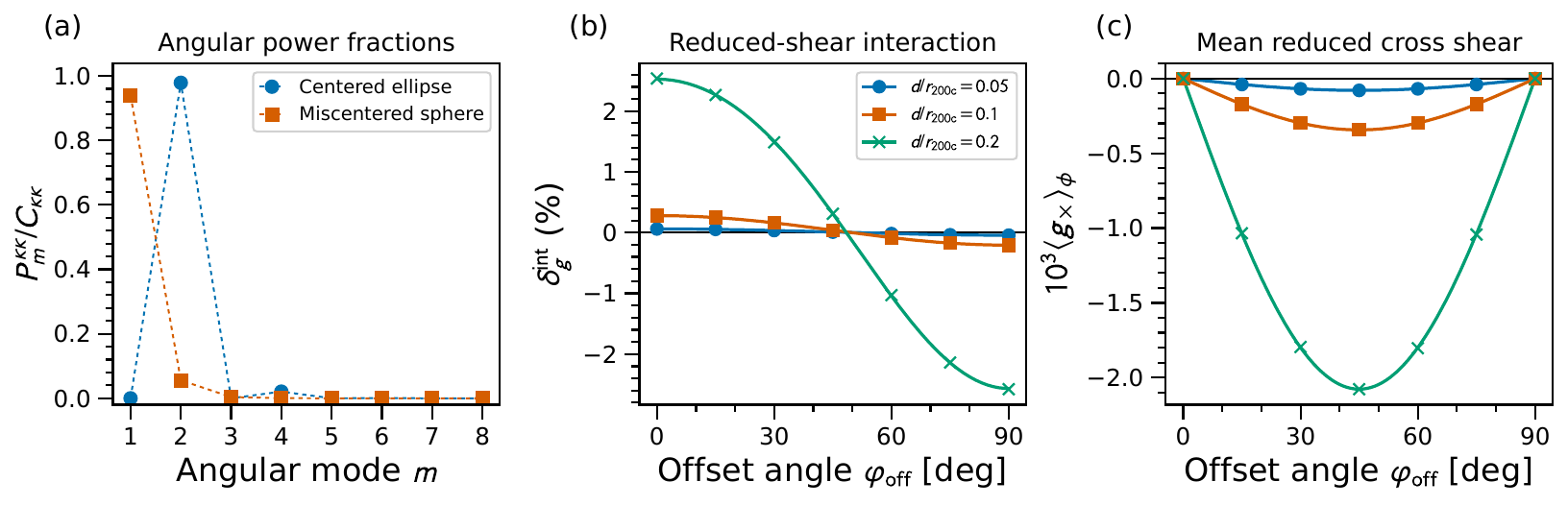}
\caption{Single-halo NFW configurations illustrating the shape--centering symmetry.  All panels use $M_{200\mathrm c}=10^{15}h^{-1}M_\odot$, $c_{200\mathrm c}=3.84$, $z_l=0.3$, and $z_s=1$.  The centering-offset angle $\phioff$ is measured from the projected major axis.  (a) Fraction $P_m^{\kappa\kappa}/\Ckk$ of the nonmonopole $\kappa$ variance carried by angular mode $m$ at $R\simeq0.3\rtwo$.  The centered elliptical halo has $q_\perp=0.67$ and $d=0$, whereas the displaced spherical halo has $q_\perp=1$ and $d/\rtwo=0.10$.  Markers show the discrete mode powers, and dashed lines connect neighboring modes only to guide the eye.  (b) Shape--centering interaction residual $\deltag^{\mathrm{int}}$ from Equation~\eqref{eq:main-interaction-exact} for displaced elliptical halos with $q_\perp=0.67$ and $d/\rtwo=0.05$, $0.10$, and $0.20$, at $R\simeq0.3\rtwo$.  Markers show the sampled values, and solid curves show fits of $A_0+A_2\cos2\phioff+A_4\cos4\phioff$. (c) For the same configurations and radius as in panel (b), we numerically average $g_\times(R,\phi)$ around the ring and show $10^3\avg{g_\times}$.  Markers show the sampled values, and solid curves show fits of $B_2\sin2\phioff$.  The same color and marker shape denote the same offset in panels (b) and (c).}
\label{fig:main-symmetry}
\end{figure*}

To examine how the leading miscentering and projected-shape perturbations combine, we consider a controlled potential containing only the corresponding $m=1$ and $m=2$ angular modes:
$\delta\psi=\delta\psi_{\mathrm{off}}^{(m=1)}+\delta\psi_{\mathrm{ell}}^{(m=2)}$.
These modes are the leading terms generated by translating the spherical reference halo and by introducing centered projected ellipticity (Figure~\ref{fig:main-ring-averaging}).  In this controlled setup, their amplitudes are taken to be linear in $u$ and $\epsp$, respectively (see Appendices~\ref{appB:potential} and \ref{appD:interaction}).
Since $\kappa$ and $\gamma$ are linear differential responses to the lensing potential, their angular fluctuations have the same modal decomposition: $\delta\bm F = \delta\bm F_{\mathrm{off}}^{(m=1)}  +\delta\bm F_{\mathrm{ell}}^{(m=2)}$. Substituting this $\delta\bm F$ into the quadratic moments $\Ckk$ and $\Ckplus$ in Equation~\eqref{eq:main-deltag-second-order} generates an $m=1\times m=1$ contribution of $O(u^2)$, an $m=2\times m=2$ contribution of $O(\epsp^2)$, and an $m=1\times m=2$ cross term of $O(\epsp u)$.  Complete-ring orthogonality removes the cross term.  

Thus, through second order in the geometric perturbations, the reduced-shear residual is additive:
\begin{equation}
 \begin{aligned}
 \deltag(R\mid\epsp,d,\phioff)
 ={}&
 \deltag^{\mathrm{ell},(2)}(R\mid\epsp)
 +\deltag^{\mathrm{off},(2)}(R\mid d)\\
 &+O(\epsp u^2,\epsp^4,u^4,\epsp^2u^2).
 \end{aligned}
 \label{eq:main-leading-additivity}
\end{equation}
Here $\deltag^{\mathrm{ell},(2)}=O(\epsp^2)$ and $\deltag^{\mathrm{off},(2)}=O(u^2)$ are obtained by evaluating Equation~\eqref{eq:main-deltag-second-order} using the isolated $m=2$ shape and $m=1$ offset fields, respectively.  
The trailing $(2)$ identifies the second-order approximation in the angular field fluctuations $\delta\bm F(R,\phi)$.  Note that the cancellation of the $m=1\times m=2$ cross term holds for each configuration and therefore does not require statistical independence between halo shape and centering offset.\footnote{Within this controlled two-mode model, the remainder is generated by higher orders in the nonlinear reduced-shear expansion.  Physical displaced triaxial halos contain additional geometric modes, which can contribute at these higher orders but do not alter the second-order $O(\epsp u)$ cancellation.}

Offset reversal, reflection, and axis interchange further constrain the first orientation-dependent term:
\begin{equation}
 \begin{aligned}
  \deltag(R\mid\epsp,d,\phioff) ={}&
  A_\epsilon(R)\epsp^2+A_u(R)u^2\\
  &+A_\varphi(R)\epsp u^2\cos2\phioff\\
  &+O(\epsp^4,u^4,\epsp^2u^2).
 \end{aligned}
 \label{eq:main-full-interaction}
\end{equation}
The $A_\epsilon$ and $A_u$ terms are the leading shape-only and offset-only contributions, respectively.  The coefficient $A_\varphi$ controls the leading relative-orientation dependence, $A_\varphi\epsp u^2\cos2\phioff$.  The dependence of the response coefficients on halo parameters and source plane is left implicit.

These two harmonic selection rules---the absence of an $O(\epsp u)$ term and the leading allowed $\epsp u^2\cos2\phioff$ orientation dependence---also govern the complete-ring residuals of smooth parity-even lensing observables, including $g_+$, $Y$, $\mu$, and $Y^q$, wherever their field expansions are regular.

For each halo realization $i$, at the same radius and source plane, we evaluate the joint shape--offset residual and the corresponding shape-only ($\mathrm{ell}$) and offset-only ($\mathrm{off}$) residuals:
\begin{equation}
 \begin{aligned}
 \delta_{g,i}
 &\equiv \deltag(R\mid\epsilon_{\perp,i},d_i,\varphi_{\mathrm{off},i}),\\
 \delta_{g,i}^{\mathrm{ell}}
 &\equiv \deltag(R\mid\epsilon_{\perp,i},d=0),\\
 \delta_{g,i}^{\mathrm{off}}
 &\equiv \deltag(R\mid\epsp=0,d_i).
 \end{aligned}
 \label{eq:main-shape-offset-exact}
\end{equation}
Unlike the second-order approximations $\deltag^{\mathrm{ell},(2)}$ and $\deltag^{\mathrm{off},(2)}$ introduced above, these residuals are calculated directly from the corresponding local lensing fields $\bm F(R,\phi)$.\footnote{For the offset-only calculation, we set $q_{\perp,i}=1$ while keeping the circular reference profile and offset vector of the projected halo unchanged (see Appendix~\ref{appC:triaxial}).}
We define their additive combination $\delta_{g,i}^{\mathrm{add}}$ and the remaining interaction $\delta_{g,i}^{\mathrm{int}}$ by
\begin{align}
 \delta_{g,i}^{\mathrm{add}}
 &\equiv
 \delta_{g,i}^{\mathrm{ell}}+\delta_{g,i}^{\mathrm{off}},
 \label{eq:main-additive-exact}\\
 \delta_{g,i}^{\mathrm{int}}
 &\equiv
 \delta_{g,i}-\delta_{g,i}^{\mathrm{add}}.
 \label{eq:main-interaction-exact}
\end{align}

For the parity-odd counterpart, the leading term is
\begin{equation}
 \avg{g_\times(R,\phi\mid\epsp,d,\phioff)}
 =B_\varphi(R)\,\epsp u^2\sin2\phioff+\cdots,
 \label{eq:main-cross-interaction}
\end{equation}
where $B_\varphi(R)$ controls the leading parity-odd relative-orientation dependence.  The $\cos2\phioff$ dependence of $\deltag^{\mathrm{int}}$ and the $\sin2\phioff$ dependence of $\avg{g_\times}$ provide complementary signatures of the same nonlinear shape--centering interaction.

Figure~\ref{fig:main-symmetry} uses the same halo mass $M_{200\mathrm c}$, NFW concentration $c_{200\mathrm c}$, and redshifts $(z_l,z_s)$ as Figure~\ref{fig:main-ring-averaging} to illustrate the shape--centering symmetry.
For the elliptical configurations in both figures, we use the projected elliptical NFW model defined in Appendix~\ref{appB:elliptical-profile}.\footnote{These illustrative configurations correspond to $b_{\mathrm{los}}=1$ in Appendix~\ref{appC:triaxial}, so that $r_{s,\perp}=r_s$ and $\kappa_{s,\perp}=\kappa_s$.  In the triaxial population experiment, both $q_\perp$ and $b_{\mathrm{los}}$ depend on intrinsic halo shape and viewing orientation, with $b_{\mathrm{los}}$ rescaling the projected NFW scale radius and normalization according to Equation~\eqref{eq:appC-projected-parameters}.}  We adopt $q_\perp=0.67$, corresponding to $\epsp\simeq0.20$.\footnote{The adopted value $q_\perp=0.67$ equals the median projected axis ratio inferred for the 20 CLASH clusters analyzed individually with two-dimensional weak lensing by \citet{Umetsu2018}.}

Panel (a) isolates the two effects by comparing a centered elliptical halo with a displaced spherical halo.  At a fixed radius, $\Ckk=\sum_{m\geq1}P_m^{\kappa\kappa}(R,R)$, where $P_m^{\kappa\kappa}$ is the contribution from angular mode $m$, as defined in Appendix~\ref{appB:potential}.  Thus, $P_m^{\kappa\kappa}/\Ckk$ is the fraction of the nonmonopole $\kappa$ variance carried by mode $m$: the centered elliptical halo is dominated by $m=2$, whereas the displaced spherical halo is dominated by $m=1$, as expected from Equation~\eqref{eq:main-m1-m2}.  Panel (b) shows that the interaction residual $\deltag^{\mathrm{int}}$ varies primarily as $\cos2\phioff$, while panel (c) shows the complementary $\sin2\phioff$ dependence of the ring-averaged reduced cross shear.

Appendix~\ref{app:coherent-halo} derives the centered-shape contribution in Equation~\eqref{eq:main-leading-additivity} from a coherent quadrupole, including its ring average and the nonlinear generation of an $m=4$ mode.  Appendix~\ref{app:miscentering-alignment} derives the spherical-offset contribution in the same equation, together with the displaced-halo fields and the symmetry transformations underlying Equations~\eqref{eq:main-full-interaction} and \eqref{eq:main-cross-interaction}.

\section{Cluster-population experiment}
\label{sec:population}

The symmetry arguments above identify how projected halo shape and centering enter the nonlinear ring-averaging residual. We now test these predictions using controlled populations of projected triaxial NFW halos, including centering offsets, paired source-redshift calculations, and a common far-background subcriticality criterion.  Table~\ref{tab:main-population-config} summarizes the numerical configuration, and Appendix~\ref{app:population-numerics} gives details of the halo projection, population model, offset sampling, and numerical accuracy checks.

\begin{deluxetable}{ll}[t]
\tabletypesize{\scriptsize}
\tablecaption{Numerical configuration of the halo-population experiment.\label{tab:main-population-config}}
\tablehead{
\colhead{Quantity} & \colhead{Value}
}
\startdata
Halo mass, $\Mtwo/(h^{-1}M_\odot)$ & $3\times10^{14},10^{15},2\times10^{15}$\\
Lens redshift, $z_l$ & $0.2,0.5$\\
Halos per population, $\Nhalo$ & $10^4$\\
Rayleigh scale of $d/\rtwo$, $\sigma_d$ & $0,0.02,0.05,0.08$\\
Source planes, $z_s$ & $1$ (fiducial), $2$ (paired test)\\
Radial grid, $R/\rtwo$ & 36 log-spaced points, $0.02$--$1$\\
Azimuthal samples, $N_\phi$ & $256$\\
Count slope, $\alpha$ & $0.3,1.4$\\
Far-background subcriticality & $\lambda_{-,\infty}\geq0.1$\\
Minimum represented fraction & $0.95$
\enddata
\tablecomments{$\sigma_d=0$ denotes the centered case; $\sigma_d=0.05$ is the reference offset model, and $\sigma_d=0.02$ and $0.08$ are narrower and broader alternatives, respectively.}
\end{deluxetable}
\tabletypesize{\small}

\subsection{Projected triaxial halos and centering offsets}
\label{subsec:main-population}

We construct the numerical populations by projecting triaxial NFW halos. Each halo is parameterized by spherical-equivalent $(\Mtwo,c_{200\mathrm c})$ variables using the $\Delta=200$ overdensity convention defined in Section~\ref{sec:intro}. We define the NFW concentration as $c_{200\mathrm c}\equiv\rtwo/r_s$, where $r_s$ is the NFW scale radius.

As summarized in Table~\ref{tab:main-population-config}, we use three halo masses, $\Mtwo/(h^{-1}M_\odot)=3\times10^{14},\ 10^{15},\ 2\times10^{15}$, at each of two lens redshifts, $z_l=0.2$ and $0.5$, giving six $(\Mtwo,z_l)$ populations. Each population contains $\Nhalo=10^4$ halo realizations. Within each population, we draw concentrations $c_{200\mathrm c}$ from a lognormal distribution with an intrinsic scatter of $0.16$~dex in $\log_{10}c_{200\mathrm c}$ about the median relation of \citet{DiemerJoyce2019}, evaluated at the corresponding $(\Mtwo,z_l)$. We draw intrinsic axis ratios from the cluster-scale model of \citet{Bonamigo2015}, evaluated at the corresponding halo peak height, and assume isotropic viewing orientations. The volume-preserving projection of each triaxial halo produces an NFW surface-density profile with area-preserving elliptical contours \citep[e.g.,][]{Umetsu2022}.

For each halo realization, we consider a centered case and three nonzero centering-offset models.  In each nonzero model, the dimensionless offset $x_d=d/\rtwo$ follows a Rayleigh distribution with scale parameter $\sigma_d=0.02$, $0.05$, or $0.08$, and the offset direction is drawn uniformly.  Here $d$ is the projected centering offset of an individual halo, whereas $\sigma_d$ sets the scale of the offset distribution in the population.  We denote the centered case by $\sigma_d=0$, adopt $\sigma_d=0.05$ as the reference offset model, and use $\sigma_d=0.02$ and $0.08$ as narrower and broader alternatives, respectively.

The reference scale $\sigma_d=0.05$ is motivated by measurements of offsets between optical and Sunyaev--Zel'dovich (SZ) centers \citep{Saro2015,Zenteno2020}.  In particular, \citet{Zenteno2020} find a Rayleigh scale of $(0.05\pm0.01)\rtwo$ for the dominant small-offset component of the separation distribution between brightest cluster galaxies (BCGs) and SZ centroids in their South Pole Telescope cluster sample.

For each halo realization, we evaluate the lensing fields and nonlinear observables at 36 logarithmically spaced radii spanning $0.02\leq R/\rtwo\leq1$, using $N_\phi=256$ azimuthal samples per ring.

The same random draws of concentration, intrinsic shape, orientation, and centering are reused in all observable and source-redshift comparisons.  The numerical experiment includes smooth host halos only; bound subhalos, correlated surrounding structure, and uncorrelated large-scale structure along the line of sight are not added explicitly.

Together, these ingredients define a controlled experiment that isolates the effects of halo shape and centering offsets.  Application to a particular survey would require halo-property distributions calibrated to its halo selection and a centering-offset model calibrated to its centering proxy, such as an X-ray, SZ, or optical center.

\subsection{Source-redshift scaling and common radial domain}
\label{subsec:main-source-redshift-scaling}

For the fiducial calculations, we use the source plane at $z_s=1$ (Table~\ref{tab:main-population-config}).  Each source redshift is treated as a separate single-plane calculation, so that no source-redshift distribution $p(z_s)$ or source-dependent weighting is included.  The paired $z_s=2$ calculation is used only to vary the lensing strength at fixed halo geometry.  For the adopted spatially flat $\Lambda$CDM cosmology, the comoving distance approaches a finite asymptotic value $\chi_\infty\equiv\lim_{z_s\to\infty}\chi(z_s)$.  We denote by $\kappa_\infty(R,\phi)$ and $\gamma_\infty(R,\phi)$ the \emph{far-background} convergence and shear corresponding to this asymptotic source-redshift limit.\footnote{We evaluate $\chi_\infty$ directly with the radiation density parameter $\Omega_{\mathrm{r}}=9.135\times10^{-5}$ and $\Omega_\Lambda=1-\Omega_{\mathrm{m}}-\Omega_{\mathrm{r}}$, rather than using a large but finite source-redshift proxy.}

With $\beta(z_l,z_s)\equiv D_{ls}/D_s$ and $\beta_\infty(z_l)\equiv\lim_{z_s\to\infty}\beta(z_l,z_s)$, we define the relative lensing efficiency $w(z_s)\equiv\beta(z_l,z_s)/\beta_\infty(z_l)$ at fixed lens redshift.  With the common $(R,\phi)$ arguments left implicit, the fields for a source plane at redshift $z_s$ are
\begin{equation}
 \kappa(z_s)=w(z_s)\kappa_\infty,
 \qquad
 \gamma(z_s)=w(z_s)\gamma_\infty.
 \label{eq:main-source-redshift-scaling}
\end{equation}
The same halo realizations are evaluated at both $z_s=1$ and $2$.  Since the projected mass distribution and centering geometry are identical realization by realization, the paired calculations rescale the lensing fields without changing the projected geometry, providing a controlled test of the dependence of the nonlinear residuals on the amplitude of the lensing fields.  The subcriticality mask defined below depends only on the common far-background fields and is thus identical for the two source planes.

The far-background fields $\kappa_\infty$ and $\gamma_\infty$ also define the deliberately conservative subcritical domain used below.  Using the eigenvalues of the lensing Jacobian defined in Equation~\eqref{eq:main-jacobian-properties}, we require
\begin{equation}
 \lambda_{-,\infty}(R)
 \equiv
 \min_\phi[1-\kappa_\infty(R,\phi)-|\gamma_\infty(R,\phi)|]
 \geq0.1.
 \label{eq:main-safety}
\end{equation}
On the sampled radial grid listed in Table~\ref{tab:main-population-config}, a realization is retained at $R_j$ only if this criterion is satisfied at $R_j$ and at every larger sampled radius, so that its retained radii form a contiguous outer domain. Since $0<w(z_s)<1$ for both source planes, the criterion provides a stronger positive-parity buffer than the corresponding condition at $z_s=1$ or $2$ and keeps every retained ring away from the far-background critical curve.

The represented fraction at $R_j$ is the fraction of realizations retained by this rule.  For the reference $\sigma_d=0.05$ model, the sampled radii with represented fraction at least $0.95$ in all six $(\Mtwo,z_l)$ populations define the common radial domain
\begin{equation}
 0.366\leq R/\rtwo\leq1.
 \label{eq:main-common-domain}
\end{equation}

\subsection{Population test of the even-power symmetry}
\label{subsec:main-shape-population}

To test the leading $O(\epsp^2)$ term in the even-power expansion of Equation~\eqref{eq:main-even-ellipticity}, we analyze the centered triaxial populations ($\sigma_d=0$) at $z_s=1$.  For consistency with the representative radius $R\simeq0.3\rtwo$ used in Figure~\ref{fig:main-symmetry}, we adopt the nearest radial-grid point, $R=0.292\rtwo$.  We retain halos satisfying the far-background subcriticality criterion, sort them by $\epsp^2$, and divide the retained halos in each $(\Mtwo,z_l)$ population into ten nearly equal-count bins in $\epsp^2$.

For each retained halo $i$, we calculate the leading response coefficient $A_{\epsilon,i}$ in Equation~\eqref{eq:main-full-interaction} from the circular projected NFW reference profile associated with that projected realization (Appendix~\ref{appC:triaxial}) at the same radius and source plane.  Combining Equations~\eqref{eq:appB-deltag-ell-second-order}, \eqref{eq:appB-density-multipoles}, and \eqref{eq:appB-shear-responses} gives
\begin{equation}
 \begin{aligned}
 A_{\epsilon}(R)
 &=
 -\frac{\kappa_{1}\left(\kappa_{1}-2\kappa_{0}+3\overline{\kappa}_{4}\right)}
 {2D_{0}\gamma_{+,0}}
 +\frac{\kappa_{1}^2}{2D_{0}^2},\\
 D_{0}&\equiv 1-\kappa_{0},
 \end{aligned}
 \label{eq:main-leading-shape-response}
\end{equation}
where $\kappa_{0}$ and $\gamma_{+,0}$ are the convergence and tangential shear of the circular projected reference profile, $\kappa_{1}\equiv d\kappa_{0}/d\ln R$, and $\overline{\kappa}_{4}$ is defined by Equation~\eqref{eq:appB-kappa4}.  The leading prediction for each halo is $A_{\epsilon,i}\epsilon_{\perp,i}^2$.  The corresponding prediction for each bin $b$ is $\operatorname{median}_{i\in b}(A_{\epsilon,i}\epsilon_{\perp,i}^2)$.

At this radius, the mask retains all halos in the three $z_l=0.2$ populations and in the $\Mtwo=3\times10^{14}h^{-1}M_\odot$, $z_l=0.5$ population.  The retained fractions are $99.9\%$ for the $\Mtwo=10^{15}h^{-1}M_\odot$, $z_l=0.5$ population and $92.6\%$ for the $\Mtwo=2\times10^{15}h^{-1}M_\odot$, $z_l=0.5$ population.\footnote{The highest-$\Mtwo$ $z_l=0.5$ sequence is shown here as a diagnostic using the retained 92.6\% of that population, whereas the accuracy results in Section~\ref{sec:accuracy} use the common 95\%-represented domain, $0.366\le R/\rtwo \le 1$.}

For each retained halo, both the exact ring average $\avg{g_+}$ and the mean-field prediction $g_{+,\mathrm{MF}}$ are constructed from the same triaxial lensing fields.  The mean-field prediction thus already includes the changes in $\avg{\kappa}$ and $\avg{\gamma_+}$ induced by triaxiality.  The residual $\deltag$ isolates the additional difference between averaging the local reduced shear and evaluating reduced shear from the azimuthally averaged fields.

\begin{figure}[t]
\centering
\includegraphics[width=0.9\columnwidth]{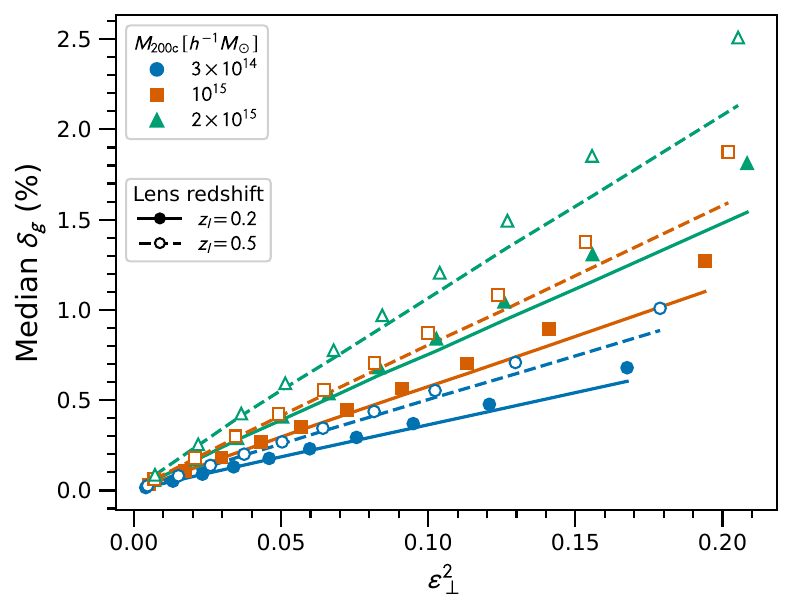}
\caption{Leading projected-shape response for centered triaxial NFW halos at $z_s=1$ and $R=0.292\rtwo$.  In each $(\Mtwo,z_l)$ population, halos satisfying the far-background subcriticality criterion are divided into ten nearly equal-count bins in $\epsp^2$.  Markers show the bin medians of $\deltag$, evaluated directly from the local lensing fields, at the corresponding bin medians of $\epsp^2$.  Lines connect the binwise medians of the halo-level leading predictions, $A_{\epsilon,i}\epsilon_{\perp,i}^2$, from Equation~\eqref{eq:main-leading-shape-response}.  Colors and marker shapes identify halo mass.  Solid lines with filled markers indicate $z_l=0.2$, and dashed lines with open markers indicate $z_l=0.5$.}
\label{fig:main-shape-closure}
\end{figure}

Figure~\ref{fig:main-shape-closure} compares the directly evaluated median residuals $\deltag$ with the corresponding binwise leading predictions.  The resulting leading predictions reproduce the mass--redshift ordering, and their relative differences from the directly evaluated medians are only $0.4$--$0.8\%$ in the lowest-$\epsp^2$ bin.  They lie progressively below the directly evaluated medians as $\epsp^2$ increases; in the highest-$\epsp^2$ bin, their relative shortfall is $11$--$15\%$ across the six populations.  This behavior is consistent with higher even-order terms beginning at $O(\epsp^4)$ becoming non-negligible toward the largest sampled $\epsp$ values.  The variation in response amplitude across mass and redshift reflects differences in lensing strength and in the radial structure of the circular projected reference profiles.

\section{How accurate is the mean-field approximation?}
\label{sec:accuracy}

Having defined the controlled halo populations and the common radial domain, we now quantify the accuracy of the mean-field approximation for the reference offset model $\sigma_d=0.05$, which is held fixed throughout this section.  In Section~\ref{subsec:main-observable-residuals}, we compare the residuals among observables. In Section~\ref{subsec:main-source-redshift-test}, we use the paired $z_s=1$ and $2$ calculations to isolate their dependence on the amplitude of the lensing fields.

\subsection{Residuals across observables}
\label{subsec:main-observable-residuals}

We calculate the residuals for reduced tangential shear $g_+$, inverse magnification $Y=\mu^{-1}$, magnification $\mu$, and magnification bias $Y^q$ from the same local lensing fields $\bm F(R,\phi)$.  The corresponding fractional residuals $\deltag$, $\deltaY$, $\deltamu$, and $\deltamb(q)$ are defined in Equations~\eqref{eq:main-deltag-definition} and \eqref{eq:main-magnification-residuals}.  The two magnification-bias examples use $\alpha=0.3$ ($q=0.7$) and $\alpha=1.4$ ($q=-0.4$), corresponding to count depletion and enhancement, respectively \citep[e.g.,][]{Umetsu2013}.

\begin{deluxetable*}{lrrrrrr}
\tabletypesize{\scriptsize}
\tablecaption{Residual quantiles at the configurations with the largest median magnitude\label{tab:main-residuals}}
\tablehead{
\colhead{} &
\multicolumn{3}{c}{$z_s=1$ (\%)} &
\multicolumn{3}{c}{$z_s=2$ (\%)} \\
\colhead{Observable} &
\colhead{$P_{16}^\ast$} & \colhead{$P_{50}^\ast$} & \colhead{$P_{84}^\ast$} &
\colhead{$P_{16}^\ast$} & \colhead{$P_{50}^\ast$} & \colhead{$P_{84}^\ast$}
}
\startdata
Reduced tangential shear, $g_+$      & +0.404  & +0.903   & +1.64   & +0.691 & +1.55   & +2.83   \\
Inverse magnification, $Y=\mu^{-1}$ & -0.105  & -0.00929 & +0.0326 & -0.262 & -0.0232 & +0.0809 \\
Magnification, $\mu$                 & +0.298  & +0.639   & +1.13   & +0.829 & +1.84   & +3.41   \\
Magnification bias, $\alpha=0.3$    & -0.155  & -0.0686  & -0.0286 & -0.417 & -0.191  & -0.0809 \\
Magnification bias, $\alpha=1.4$    & +0.0845 & +0.179   & +0.315  & +0.232 & +0.507  & +0.924
\enddata
\tablecomments{For each observable and source redshift, the selected configuration maximizes $\left|P_{50}(M_{200\mathrm c},z_l,R/\rtwo)\right|$ over the six $(\Mtwo,z_l)$ grid points and the sampled scaled radii $R/\rtwo$ in the common radial domain.  For the reference offset model $\sigma_d=0.05$, at least 95\% of every $(\Mtwo,z_l)$ population satisfies the far-background subcriticality criterion throughout this domain.  The quoted $P_{16}^\ast$, $P_{50}^\ast$, and $P_{84}^\ast$ are evaluated from the retained halos at the selected configuration, with $P_{16}^\ast$--$P_{84}^\ast$ characterizing the halo-to-halo variation.  The selection is performed independently for each observable and source redshift; all combinations select $\Mtwo=2\times10^{15}h^{-1}M_\odot$, $z_l=0.5$, and $R/\rtwo=0.366$.  The $z_s=1$ and $2$ calculations use the same halo and centering realizations and the same far-background subcriticality mask.}
\end{deluxetable*}

Each fractional residual compares the exact ring average of the locally evaluated observable with the mean-field prediction formed by applying the same nonlinear map to the ring-averaged fields. Both are constructed from the same local fields at the same radius, adopted center, and source plane. Thus, changes in $\avg{\kappa}$ and $\avg{\gamma_+}$ caused by triaxiality or miscentering enter both terms; the residual isolates the additional difference between the two orders of operation.

For each observable $\mathcal{O}$ and source redshift $z_s$, we evaluate the fractional residual $\delta_{\mathcal{O}}$ at each of the six $(\Mtwo,z_l)$ grid points and each sampled scaled radius $R/\rtwo$ in the common radial domain, using every halo that satisfies the far-background subcriticality criterion.  By construction, at least 95\% of each $(\Mtwo,z_l)$ population is retained throughout this domain.  At fixed $(M_{200\mathrm c},z_l,R/\rtwo)$, the halo-to-halo distribution of $\delta_{\mathcal{O}}$ is generated by the Monte Carlo variations in concentration, intrinsic triaxial shape, viewing orientation, and centering-offset amplitude and direction.  No observational noise is included.

For fixed $\mathcal{O}$ and $z_s$, let $P_p(M_{200\mathrm c},z_l,R/\rtwo)$ denote the $p$th percentile of this halo-to-halo residual distribution.  In particular, $P_{50}$ is the population median at that halo mass, lens redshift, and scaled radius.

For each observable and source redshift, we then identify the configuration at which the magnitude of the population median is largest over the six $(\Mtwo,z_l)$ grid points and all sampled scaled radii $R/\rtwo$ in the common radial domain.  Denoting this configuration by $\left(M_{200\mathrm c,\ast},z_{l,\ast},(R/\rtwo)_\ast\right)$, we define
\begin{equation}
P_p^\ast
\equiv
P_p\left(M_{200\mathrm c,\ast},z_{l,\ast},(R/\rtwo)_\ast\right).
\label{eq:main-extremal-quantiles}
\end{equation}
The selection is performed independently for each observable and source redshift.  Table~\ref{tab:main-residuals} reports $P_{16}^\ast$, $P_{50}^\ast$, and $P_{84}^\ast$.  Here $P_{50}^\ast$ retains the sign of the population median at the selected extremal configuration, while $P_{16}^\ast$ and $P_{84}^\ast$ characterize the halo-to-halo spread at that same configuration.  Despite the independent selection, all observables at both source redshifts select $\Mtwo=2\times10^{15}h^{-1}M_\odot$, $z_l=0.5$, and $R/\rtwo=0.366$.

\begin{figure}[t]
\centering
\includegraphics[width=0.9\columnwidth]{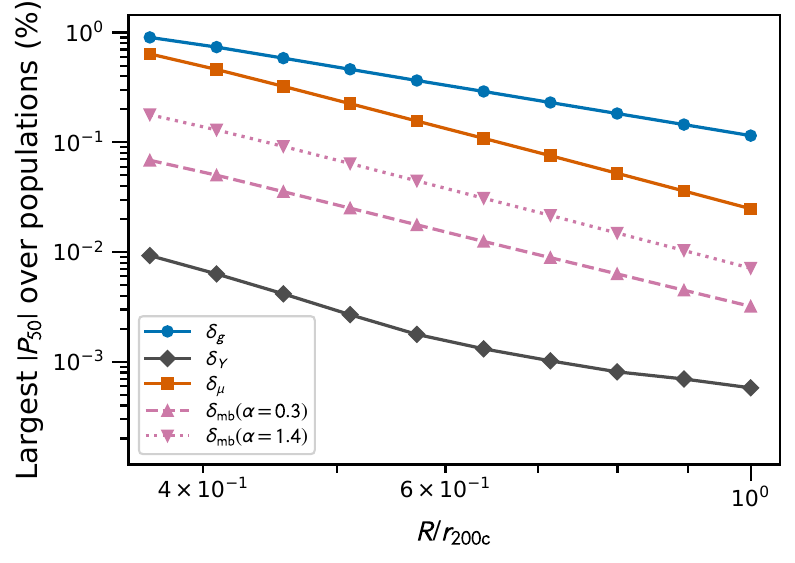}
\caption{Radial dependence of the median residuals for the reference $\sigma_d=0.05$ populations at $z_s=1$.  At each sampled scaled radius $R/\rtwo$, $P_{50}$ is calculated separately for the six $(\Mtwo,z_l)$ populations using halos that satisfy the far-background subcriticality criterion, and the largest $|P_{50}|$ is shown.  Only the common radial domain, where at least 95\% of the halos in every population are retained, is included.  The two magnification-bias cases share a common color, while marker shape and line style distinguish $\alpha=0.3$ and $1.4$.}
\label{fig:main-envelope}
\end{figure}

Figure~\ref{fig:main-envelope} complements Table~\ref{tab:main-residuals} by retaining the radial dependence of the population medians for $z_s=1$.  At each sampled $R/\rtwo$, each curve gives the largest $|P_{50}|$ among the six $(\Mtwo,z_l)$ populations; the maximum along each curve thus gives $|P_{50}^\ast|$ in Table~\ref{tab:main-residuals}.  Since all observables are evaluated for the same halo and centering realizations, differences among the curves reflect their different nonlinear dependence on the lensing fields.

Inverse magnification shows a qualitatively different population behavior.  For the fiducial $z_s=1$ calculation, its extremal median is $P_{50}^\ast=-0.00929\%$, while the corresponding $P_{16}^\ast$--$P_{84}^\ast$ interval spans both signs.  Equation~\eqref{eq:main-Y-exact} gives $\avg{Y}-\YMf=\Ckk-\Cplusplus-\Ccrosscross$, involving only quadratic combinations of the local lensing fields.  The small median reflects partial cancellation between the $\kappa$ and $\gamma$ contributions to this combination, while the population quantiles reveal appreciable halo-to-halo variation around a median close to zero.

Magnification is more sensitive because the nonlinear inversion $\mu=Y^{-1}$ amplifies angular fluctuations as $Y$ decreases.  The two magnification-bias responses remain smaller in magnitude than magnification itself over the common radial domain, as seen in Figure~\ref{fig:main-envelope}.

\subsection{Dependence on source redshift}
\label{subsec:main-source-redshift-test}

The paired $z_s=1$ and $2$ calculations provide a controlled test of the dependence of the nonlinear residuals on the amplitude of the lensing fields. The projected mass distribution, centering geometry, and far-background subcriticality mask are held fixed realization by realization; changing $z_s$ only rescales the convergence and shear according to Equation~\eqref{eq:main-source-redshift-scaling}.  As shown in Table~\ref{tab:main-residuals}, this increase in lensing amplitude increases $|P_{50}^\ast|$ for every observable.

\begin{figure}[t]
\centering
\includegraphics[width=0.9\columnwidth]{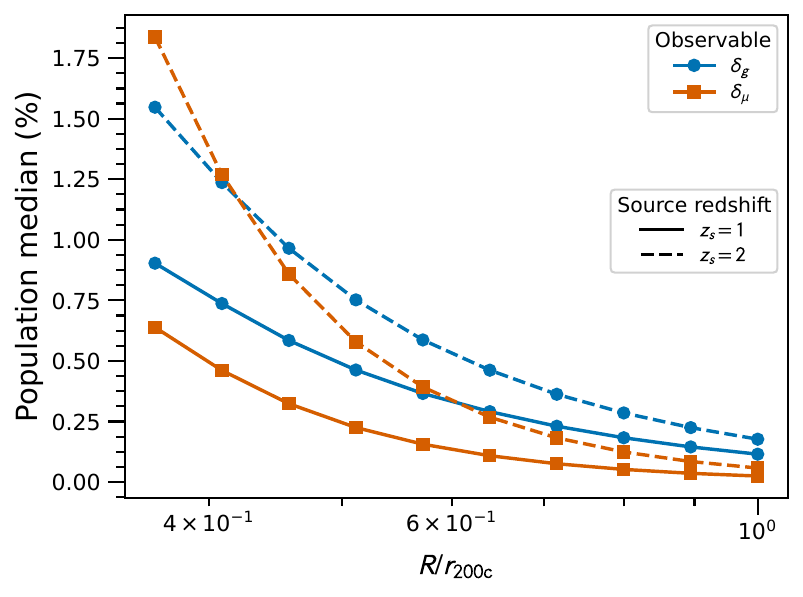}
\caption{Source-redshift dependence of the median residuals for the $\Mtwo=2\times10^{15}h^{-1}M_\odot$, $z_l=0.5$, and $\sigma_d=0.05$ population.  At each sampled scaled radius $R/\rtwo$, $P_{50}$ is evaluated for reduced tangential shear and magnification using halos that satisfy the far-background subcriticality criterion.  The same retained halo realizations are used at $z_s=1$ and $2$.  Color and marker shape distinguish the two observables, while solid and dashed lines denote $z_s=1$ and $2$, respectively. Only the common radial domain defined for the reference $\sigma_d=0.05$ populations is shown.}
\label{fig:main-source-redshift}
\end{figure}

Figure~\ref{fig:main-source-redshift} retains the radial dependence for the population with $\Mtwo=2\times10^{15}h^{-1}M_\odot$ and $z_l=0.5$, the $(\Mtwo,z_l)$ grid point that sets the extremal medians in Table~\ref{tab:main-residuals}.
For clarity, the figure shows reduced tangential shear and magnification, which have the largest $|P_{50}^\ast|$ among the observables considered.  Their larger residuals at $z_s=2$ thus directly trace the response to the increased lensing amplitude rather than a change in halo structure or in the retained population.

\section{Accuracy limits, conditioning, and robustness}
\label{sec:limits}

Section~\ref{sec:accuracy} quantified the baseline accuracy for all five observables across the six $(\Mtwo,z_l)$ populations with the reference offset model $\sigma_d=0.05$ over the common radial domain, where all six populations are at least 95\% represented.  We now relax these baseline restrictions as needed in controlled diagnostic tests, while retaining the far-background subcriticality criterion.

The diagnostics below focus on reduced tangential shear.  Section~\ref{subsec:main-large-u} tests the shape--centering interaction model developed in Section~\ref{subsec:main-interaction} as $u=d/R$ increases.  Section~\ref{subsec:main-conditioning} examines the amplification of $\deltag$ when $g_{+,\mathrm{MF}}$ is close to zero.  Section~\ref{subsec:main-criticality} distinguishes the formal pole of the reduced-shear parametrization at $1-\kappa=0$ from lensing criticality at $Y=0$.  Finally, Section~\ref{subsec:main-alignment} tests the robustness of $\deltag$ and its hybrid predictor to correlations between centering-offset directions and projected halo shape.

\subsection{Validity of the reduced-shear shape--centering interaction model}
\label{subsec:main-large-u}

Equation~\eqref{eq:main-full-interaction} identifies $u=d/R$ as the local translation variable and predicts that the leading orientation-dependent shape--centering interaction scales as $\epsp u^2\cos2\phioff$.  

Here, we test the range over which this leading interaction term, combined with the shape-only and offset-only residuals, describes the reduced-shear residual $\deltag$.  We use the $z_s=1$ calculations for all six $(\Mtwo,z_l)$ grid points and the three nonzero offset models, $\sigma_d=0.02$, $0.05$, and $0.08$.  We fit each $(\Mtwo,z_l,R/\rtwo,\sigma_d)$ configuration separately.  Together, the three offset scales test whether $u=d/R$ provides a common validity criterion across the different offset distributions.  We evaluate four sampled scaled radii,
\begin{equation}
 R/\rtwo=0.209,\ 0.292,\ 0.511,\ 1.000.
 \label{eq:main-closure-radii}
\end{equation}
These radii include values outside the common radial domain used for the baseline calculation in Section~\ref{sec:accuracy}.  At each radius, we use the halos that satisfy the far-background subcriticality criterion.  The six grid points combined with the three offset models define 18 samples at each radius, giving $4\times18=72$ samples in total.

For a fixed $(\Mtwo,z_l,R/\rtwo,\sigma_d)$ configuration, the sample consists of the retained Monte Carlo halo realizations, which vary in concentration, intrinsic axis ratios, viewing orientation, and centering-offset amplitude and direction.  We denote the reduced-shear residual of retained halo $i$ by
\begin{equation}
 y_i\equiv\delta_{g,i}.
 \label{eq:main-closure-target}
\end{equation}
Using the additive combination $\delta_{g,i}^{\mathrm{add}}$ of the shape-only and offset-only residuals defined in Equation~\eqref{eq:main-additive-exact}, we form the predictor
\begin{equation}
 \widehat y_i
 =\delta_{g,i}^{\mathrm{add}}
 +a_0
 +a_{\mathrm{int}}\,\epsilon_{\perp,i}u_i^2
 \cos2\varphi_{\mathrm{off},i}.
 \label{eq:main-closure-predictor}
\end{equation}
Equation~\eqref{eq:main-closure-predictor} is a hybrid residual predictor: $\delta_{g,i}^{\mathrm{add}}$ contains the pure-shape and pure-offset residuals including all orders in those separate contributions.  Only the remaining nonadditive interaction is modeled by the intercept and the leading orientation-dependent term.  The coefficient of $\delta_{g,i}^{\mathrm{add}}$ is fixed to unity.  For each sample, $a_0$ and $a_{\mathrm{int}}$ are determined by unweighted least squares.  The intercept $a_0$ absorbs the sample-mean interaction remainder not represented by the leading $\cos2\varphi_{\mathrm{off}}$ term.

We quantify the accuracy of Equation~\eqref{eq:main-closure-predictor} using the coefficient of determination,
\begin{equation}
 \mathcal{R}^2
 \equiv
 1-\frac{\sum_{i=1}^{N}(y_i-\widehat y_i)^2}
         {\sum_{i=1}^{N}(y_i-y_{\mathrm{mean}})^2},
 \label{eq:main-R2-definition}
\end{equation}
where $y_{\mathrm{mean}} \equiv (1/N)\sum_{i=1}^{N}y_i$ is the sample mean across the retained halos.
Thus $\mathcal{R}^2=1$ corresponds to perfect prediction, while $\mathcal{R}^2=0$ gives the same total squared error as the constant predictor $\widehat y_i=y_{\mathrm{mean}}$. Negative values indicate that the hybrid predictor performs worse than this constant baseline. The coefficient $\mathcal{R}^2$ hence measures the accuracy of the hybrid reconstruction.

To test the regime in which this low-order interaction model remains accurate, we repeat the fits using only halos that satisfy
\begin{equation}
 u_i=\frac{d_i}{R}<0.3.
 \label{eq:main-u-cut}
\end{equation}
With this restriction, the predictor reproduces the residuals of individual halos extremely well across all 72 samples: $\mathcal{R}^2_{\min}=0.9944$ and $\mathcal{R}^2_{\mathrm{med}}=0.9989$.

The effect of the $u<0.3$ restriction is largest at small radius $R$, where a given offset $d$ corresponds to a larger value of $u=d/R$.  At $R=0.209\rtwo$, the minimum and median $\mathcal{R}^2$ values across the 18 $(\Mtwo,z_l,\sigma_d)$ combinations increase from $\mathcal{R}^2_{\min}=-0.783$ and $\mathcal{R}^2_{\mathrm{med}}=0.577$ without the restriction to $0.9944$ and $0.9982$ with it.  At $R=\rtwo$, the same offset distributions correspond to much smaller $u=d/R$, and even the unrestricted samples give $\mathcal{R}^2_{\min}=0.9988$ and $\mathcal{R}^2_{\mathrm{med}}=0.9999$.

This radial behavior, together with the uniformly high $\mathcal{R}^2$ values obtained after applying the same $u<0.3$ restriction, supports $u=d/R$ as the variable controlling the breakdown of the low-order shape--centering interaction model.  For the populations tested here, $u<0.3$ defines an empirically validated regime in which Equation~\eqref{eq:main-closure-predictor} accurately reconstructs the reduced-shear residual $\deltag$.

\subsection{Conditioning of the reduced-shear fractional residual}
\label{subsec:main-conditioning}

The fractional residual $\deltag$ can have a large magnitude when the difference $\Delta_\phi g_+$ between the ring average and the mean-field prediction is large, when the mean-field signal $g_{+,\mathrm{MF}}$ is close to zero, or both.

For a displaced halo realization, leaving its other parameters implicit, we write
\begin{equation}
 \begin{aligned}
 \Delta_\phi g_+(R\mid d)
 &\equiv
 \avg{g_+(R,\phi\mid d)}
 -g_{+,\mathrm{MF}}(R\mid d),\\
 \deltag(R\mid d)
 &=
 \frac{\Delta_\phi g_+(R\mid d)}
 {g_{+,\mathrm{MF}}(R\mid d)}.
 \end{aligned}
 \label{eq:main-deltag-conditioning}
\end{equation}

Using Equation~\eqref{eq:main-tangential-shear-monopole}, the mean-field signal for the same realization is
\begin{equation}
 g_{+,\mathrm{MF}}(R\mid d)
 =
 \frac{\overline{\kappa}(<R\mid d)
       -\avg{\kappa(R,\phi\mid d)}}
 {1-\avg{\kappa(R,\phi\mid d)}},
 \label{eq:main-mf-monopole}
\end{equation}
where $\overline{\kappa}(<R\mid d)$ is the interior mean defined in Equation~\eqref{eq:main-interior-mean-convergence}, evaluated about the same adopted center.  Since $1-\avg{\kappa}>0$ under the subcritical restriction, a zero of $g_{+,\mathrm{MF}}$ is a zero of the azimuthally averaged excess convergence.  At such a zero, $\deltag$ is undefined even though $\Delta_\phi g_+$ and the local lensing fields $\bm F(R,\phi)$ can remain finite.

Displaced configurations with $g_{+,\mathrm{MF}}$ close to zero can thus generate long tails in $\deltag$ even when $\Delta_\phi g_+$ remains finite.  This ratio conditioning is most relevant at small radius for the broader centering-offset distributions tested here.  Throughout this subsection, the comparisons based on $\deltag$ and $\Delta_\phi g_+$ include all halos satisfying the far-background subcriticality criterion at each radius.

To diagnose this effect, we compare the directly evaluated fractional residual $\delta_{g,i}$ with its second-order field-expansion prediction $\delta_{g,i}^{(2)}$ and separately compare the corresponding difference $\Delta_\phi g_{+,i}$ with its prediction $\Delta_\phi g_{+,i}^{(2)}$. All four quantities are constructed from the same displaced triaxial lensing fields. For realizations with $g_{+,\mathrm{MF},i}\neq0$, they satisfy
\begin{equation}
 \begin{aligned}
 \Delta_\phi g_{+,i}
 &=
 g_{+,\mathrm{MF},i}\delta_{g,i},\\
 \Delta_\phi g_{+,i}^{(2)}
 &=
 g_{+,\mathrm{MF},i}\delta_{g,i}^{(2)}
 =
 \frac{C_{\kappa+,i}}{D_i^2}
 +\frac{\avg{\gamma_{+,i}}C_{\kappa\kappa,i}}{D_i^3},
 \end{aligned}
 \label{eq:main-conditioning-predictors}
\end{equation}
where $D_i\equiv1-\avg{\kappa_i}$.

We evaluate the coefficient of determination $\mathcal{R}^2$ defined in Equation~\eqref{eq:main-R2-definition} separately for
\begin{equation}
 (y_i,\widehat y_i)
 =
 \begin{cases}
 (\delta_{g,i},\delta_{g,i}^{(2)}),
 & \text{fractional comparison},\\
 (\Delta_\phi g_{+,i},\Delta_\phi g_{+,i}^{(2)}),
 & \text{difference comparison}.
 \end{cases}
 \label{eq:main-conditioning-R2-pairs}
\end{equation}
For each comparison, $y_{\mathrm{mean}}$ in Equation~\eqref{eq:main-R2-definition} is recomputed from the corresponding target variable. At the level of individual realizations, dividing $\Delta_\phi g_{+,i}$ and $\Delta_\phi g_{+,i}^{(2)}$ by the same $g_{+,\mathrm{MF},i}$ gives $\delta_{g,i}$ and $\delta_{g,i}^{(2)}$, respectively. Since this divisor varies across the sample, the resulting $\mathcal{R}^2$ values need not agree.

To expose the conditioning effect, we use the $z_s=1$, $\Mtwo=3\times10^{14}h^{-1}M_\odot$, $z_l=0.2$ population with the largest tested centering-offset scale, $\sigma_d=0.08$, at $R=0.209\rtwo$.  The fractional and difference comparisons give $\mathcal{R}^2=0.901$ and $0.948$, respectively.  The higher value for the difference comparison shows that division by realization-dependent values of $g_{+,\mathrm{MF}}$ close to zero contributes to the lower $\mathcal{R}^2$ in the fractional comparison.  The remaining departure from unity shows that terms beyond second order in the field expansion also contribute at this radius.  At $R=0.511\rtwo$, both comparisons give $\mathcal{R}^2\simeq0.999$.

Ratio conditioning is distinct from the breakdown of the low-order shape--centering interaction model at large $u$ examined in Section~\ref{subsec:main-large-u} and from the nonlinear boundaries considered in Section~\ref{subsec:main-criticality}.

\subsection{Reduced-shear denominators and magnification criticality}
\label{subsec:main-criticality}

Our subcritical restriction $\lambda_->0$ ensures $1-\kappa>0$ and $Y>0$ throughout each ring.  The stronger far-background criterion in Equation~\eqref{eq:main-safety} keeps the numerical sample away from the boundaries $1-\kappa=0$ and $Y=0$.

For reduced shear, let $D(R)\equiv1-\avg{\kappa(R,\phi)}$.  The local denominator is then $1-\kappa(R,\phi)=D(R)-\delta\kappa(R,\phi)$.  The Taylor expansion about the azimuthally averaged fields requires $|\delta\kappa(R,\phi)|<|D(R)|$ at every azimuth $\phi$ on the ring.

For the coherent-quadrupole fields defined in Equation~\eqref{eq:appB-quadrupole-fields}, the reduced tangential shear is
\begin{equation}
 g_+(R,\phi)
 =
 \frac{\gpavg+F_2^{+}\cos2\phi}
 {D-F_2^{\kappa}\cos2\phi},
\end{equation}
where $F_2^\kappa(R)$ and $F_2^+(R)$ are the $m=2$ field amplitudes defined in Equation~\eqref{eq:main-m2-amplitudes}.

The denominator $D-F_2^\kappa\cos2\phi$ remains positive at every azimuth if and only if $D>|F_2^\kappa|$. Under this condition, Appendix~\ref{appB:quadrupole} derives the exact ring average, the associated residual $\Delta_\phi g_+$, and the perturbative expansion of that residual in Equations~\eqref{eq:appB-exact-g}--\eqref{eq:appB-quadrupole-expansion}.  The geometric expansion converges uniformly around the ring for each fixed $|F_2^\kappa|/D<1$, while convergence slows as this ratio approaches unity.  The zero-denominator boundary $D=|F_2^\kappa|$ lies outside the strictly subcritical domain and is excluded by the numerical mask.

Critical curves occur where the determinant of the lensing Jacobian defined in Equation~\eqref{eq:main-lensing-jacobian} vanishes,
\begin{equation}
 \det\mathcal{A}(R,\phi)=Y(R,\phi)=0.
 \label{eq:main-magnification-criticality}
\end{equation}
Thus, $1-\kappa=0$ is a formal pole of the reduced-shear parametrization, whereas $Y=0$ marks lensing criticality.  As $Y\to0^+$, inverse powers diverge, while positive noninteger powers remain finite but can have singular derivatives.

The threshold $\lambda_{-,\infty}=0.1$ provides a numerical safety buffer and should not be interpreted as a universal boundary for the accuracy or breakdown of the mean-field approximation.

\subsection{Effect of offset--major-axis alignment on the reduced-shear residual}
\label{subsec:main-alignment}

Hydrodynamical simulations indicate that centering-offset directions need not be random relative to projected cluster structure \citep{Sommer2024,Sommer2025}.  The sign and magnitude of the offset--major-axis alignment statistic defined below remain unconstrained by these studies.  We thus explore both major- and minor-axis preferences over a symmetric diagnostic range.

In this subsection, we examine the population median of the reduced-shear residual $\deltag$ and the accuracy of the hybrid predictor defined in Section~\ref{subsec:main-large-u}.

We characterize the leading $m=2$ alignment between the centering-offset direction and the projected halo major axis by
\begin{equation}
 \Qoff\equiv\langle\cos2\phioff\rangle,
 \label{eq:main-Qoff}
\end{equation}
where $\phioff$ is the polar angle of the centering-offset vector $\bm d$, measured from the projected major axis, and the brackets denote an average over the population distribution of $\phioff$.  Positive $\Qoff$ favors offsets along the projected major axis, whereas negative $\Qoff$ favors the minor axis.  The uniformly distributed offset directions in our baseline populations correspond to $\Qoff=0$ in the parent distribution, for which the leading orientation-dependent interaction in Equation~\eqref{eq:main-full-interaction} averages to zero.

Since our baseline populations sample $\phioff$ uniformly, we construct aligned populations by importance reweighting their Monte Carlo realizations.  The reweighting preserves the concentrations, intrinsic shapes, viewing orientations, offset amplitudes, and lensing fields of the individual halos while changing their statistical weights according to $\phioff$.

Throughout this subsection, we evaluate each halo population using its far-background lensing fields.  The calculation covers all six $(\Mtwo,z_l)$ grid points, the three nonzero centering-offset models $\sigma_d=0.02$, $0.05$, and $0.08$, and the four scaled radii $R/\rtwo$ in Equation~\eqref{eq:main-closure-radii}.  We evaluate $\Qoff=-0.30,-0.15,0,0.15,$ and $0.30$.

For our primary reweighting model, the target angular distribution is
\begin{equation}
 p_{\mathrm{lin}}(\phioff\mid\Qoff)
 =\frac{1+2\Qoff\cos2\phioff}{2\pi},
\end{equation}
where nonnegativity requires $|\Qoff|\leq1/2$.\footnote{This restriction is specific to $p_{\mathrm{lin}}$; for a general angular distribution, the moment $\Qoff$ lies in $[-1,1]$.} The model is linear in $\Qoff$ and satisfies $\langle\cos2\phioff\rangle=\Qoff$. We thus refer to it as the linear-in-$\Qoff$ angular model. It represents the leading reflection-symmetric quadrupolar departure from isotropy and has vanishing higher Fourier moments.

For comparison, we also use a von Mises angular model with the same value of $\Qoff$ and, for nonzero $\Qoff$, nonzero higher Fourier moments (Appendix~\ref{appD:alignment}).

Using the linear-in-$\Qoff$ angular model, we first quantify how alignment changes the population median of $\deltag$ at $R=0.292\rtwo$, the representative grid radius used in Section~\ref{subsec:main-shape-population}, for the reference offset model $\sigma_d=0.05$.  Four of the six $(\Mtwo,z_l)$ populations have a weighted represented fraction of at least $95\%$ throughout $-0.30\leq\Qoff\leq0.30$.  Table~\ref{tab:main-alignment-response} gives their weighted median $\deltag$ values.

\begin{deluxetable}{ccccc}
\tabletypesize{\scriptsize}
\tablecaption{Offset--major-axis alignment response
\label{tab:main-alignment-response}}
\tablehead{
\colhead{$\Mtwo/(10^{14}h^{-1}M_\odot)$} &
\colhead{$z_l$} &
\multicolumn{3}{c}{Weighted median $\deltag$ (\%)}\\
\cline{3-5}
\colhead{} &
\colhead{} &
\colhead{$\Qoff=-0.30$} &
\colhead{$0$} &
\colhead{$+0.30$}
}
\startdata
3  & 0.2 & 0.448 & 0.466 & 0.486\\
10 & 0.2 & 0.821 & 0.857 & 0.895\\
20 & 0.2 & 1.256 & 1.301 & 1.361\\
3  & 0.5 & 1.247 & 1.306 & 1.384
\enddata
\tablecomments{Values are evaluated at $R=0.292\rtwo$ and $\sigma_d=0.05$ using the far-background lensing fields and the linear-in-$\Qoff$ angular model defined in Section~\ref{subsec:main-alignment}.  Only populations whose weighted represented fraction remains at least 95\% over $-0.30\leq\Qoff\leq0.30$ are shown; the two higher-mass $z_l=0.5$ populations do not satisfy this requirement at this radius.}
\end{deluxetable}

In all four populations, the weighted median $\deltag$ is positive.  Major-axis alignment increases its magnitude and thus slightly reduces the accuracy of the mean-field approximation as measured by this population statistic. In contrast, minor-axis alignment has the opposite effect, consistent with the $\cos2\phioff$ dependence in Equation~\eqref{eq:main-full-interaction}.

We next compare the effects of alignment and centering-offset scale for these four populations at $R=0.292\rtwo$.  Changing $\Qoff$ from $0$ to $+0.30$ at $\sigma_d=0.05$ increases the weighted median $\deltag$ by $0.020$--$0.078$ percentage points, whereas changing $\sigma_d$ from $0.02$ to $0.08$ at $\Qoff=0$ changes it by $0.380$--$1.069$ percentage points.  Thus, for the parameter changes compared here, varying the centering-offset scale produces a larger change in the weighted median $\deltag$ than introducing the tested major-axis alignment.

We also test whether the alignment response depends on the form of the angular distribution $p(\phioff\mid\Qoff)$. The two angular models coincide at $\Qoff=0$; at fixed nonzero $\Qoff$, they share the same $m=2$ moment but differ in their higher Fourier moments. Across all $(\Mtwo,z_l,\sigma_d,R/\rtwo)$ configurations satisfying the $95\%$ representation requirement throughout the tested $\Qoff$ range, their weighted median $\deltag$ values differ by at most $0.0056$ percentage points. Thus, for the two models tested, the shared $m=2$ moment $\Qoff$ largely captures the dependence of the weighted median $\deltag$ on the form of the angular distribution, with little sensitivity to the higher Fourier moments that distinguish the models.

Finally, at each $(\Mtwo,z_l,\sigma_d,R/\rtwo)$ configuration, we fit $a_0$ and $a_{\mathrm{int}}$ in Equation~\eqref{eq:main-closure-predictor} using the retained $u<0.3$ subset of the random-direction population with $\Qoff=0$.  We keep these coefficients fixed while reweighting the same realizations to nonzero $\Qoff$.  The resulting $\mathcal{R}_w^2$, defined in Equation~\eqref{eq:appD-weighted-R2}, measures how well the predictor fitted to random offset directions describes aligned populations at the same halo mass, lens redshift, offset scale, and scaled radius.

For the linear-in-$\Qoff$ angular model, the predictor fitted to the random-direction population remains accurate for aligned populations within the empirically validated regime $u<0.3$: every tested alignment sample containing at least 100 retained halos has $\mathcal{R}_w^2\geq0.9886$, and the median $\mathcal{R}_w^2$ across these samples is $0.9983$.

\section{Discussion}
\label{sec:discussion}

\subsection{Why does the mean-field approximation work?}
\label{subsec:discussion-why}

Three mechanisms suppress the fractional residual arising from the noncommutativity in Equation~\eqref{eq:main-noncommutativity}.

First, the mean-field prediction is the zeroth-order term in the expansion about $\avg{\bm F}$. Complete-ring averaging eliminates the linear term, so that the residual begins with the Hessian--covariance contraction in Equation~\eqref{eq:main-residual-hierarchy}; higher central moments enter at higher orders. This result is independent of the physical origin of $\delta\bm F$.

Second, the coherent shape and centering perturbations considered here obey additional rotational and harmonic constraints. For a centered elliptical halo, rotational symmetry makes the ring-averaged scalar residual even in signed projected ellipticity $\epsp$. Miscentering generates leading $m=1$ structure, whereas projected shape begins at $m=2$; complete-ring orthogonality thus removes the $O(\epsp u)$ interaction, where $u=d/R$. The first orientation-dependent interaction for reduced tangential shear appears at $O(\epsp u^2\cos2\phioff)$, with the complementary reduced-cross response at $O(\epsp u^2\sin2\phioff)$.

Third, the weak-lensing regime provides an additional suppression through the amplitude of the lensing fields. If the derivatives of the observable remain regular, the generic residual $\Delta_\phi\mathcal{O}$ is $O(\eta^2\epsilon^2)$. For observables with $\mathcal{O}_{\mathrm{MF}}=O(\eta)$, such as reduced shear, the corresponding fractional residual is $O(\eta\epsilon^2)$. The paired $z_s=1$ and $2$ calculations show a dependence consistent with this power counting: for every observable, rescaling the lensing fields upward at fixed halo geometry increases the largest magnitude of the population median of the fractional residual.

Together, these mechanisms explain why, for the reference offset model $\sigma_d=0.05$ at $z_s=1$, the population median of the fractional residual remains below $1\%$ in magnitude for all five fiducial observables throughout the common radial domain. The differences among observables are discussed next.

\subsection{Observable dependence}
\label{subsec:discussion-observables}

Since all observables were evaluated from the same lensing fields for each halo realization, differences among their residuals reflect the different nonlinear maps applied to those fields. For reduced tangential shear, the $(\kappa,\gamma_+)$ block of the Hessian is indefinite (Equation~\eqref{eq:appA-g-hessian}). Its leading correction combines the covariance $\Ckplus=\avg{\delta\kappa\,\delta\gamma_+}$ with the convergence variance $\Ckk=\avg{\delta\kappa^2}$ (Equation~\eqref{eq:main-delta-g}) and therefore has no universal sign.

Inverse magnification behaves differently because $Y=\mu^{-1}$ is quadratic, with positive curvature in $\kappa$ and negative curvature in the two shear components. Its residual, $\Delta_\phi Y=\Ckk-\Cplusplus-\Ccrosscross$, contains no higher field orders (Equation~\eqref{eq:main-Y-exact}). In the fiducial populations, the population median of the fractional residual $\deltaY$ for $Y$ is particularly small, in part because these three variance contributions cancel numerically; this cancellation is population dependent rather than a consequence of quadraticity. 

Magnification and magnification bias apply the additional map $Y\mapsto Y^q$, with $q=-1$ for magnification. At second order, Equation~\eqref{eq:main-Yq-second-order} separates their fractional residual into a term proportional to $\Delta_\phi Y$ and a variance term whose coefficient $q(q-1)/2$ is set by the curvature of the power-law map. This curvature fixes the sign of the variance term, whereas the total residual $\deltamb(q)$ has no universal sign because $\Delta_\phi Y$ can be positive or negative.  At all orders, the factorization in Equation~\eqref{eq:main-Yq-factorization} separates the shift from $\YMf$ to $\avg{Y}$ from the additional effect of applying $Y\mapsto Y^q$ before ring averaging. These differences among the nonlinear maps explain the observable-dependent amplitudes summarized in Table~\ref{tab:main-residuals}.

The cross component provides a related comparison between a linear field and its reduced counterpart. For a scalar lensing potential, $\avg{\gamma_\times}$ vanishes identically on any complete ring (Equation~\eqref{eq:main-true-cross-zero}), whereas the fluctuating denominator in $g_\times=\gamma_\times/(1-\kappa)$ introduces the covariance $\Ckcross$, allowing $\avg{g_\times}$ to be nonzero for an individual non-axisymmetric realization (Equation~\eqref{eq:main-reduced-cross}).  For a parity-invariant population, in which every configuration and its mirror image have equal statistical weight, the ensemble expectation of $\avg{g_\times}$ still vanishes.  The shape--centering calculation provides a concrete example: an individual translated elliptical halo can have a nonzero $\avg{g_\times}$ whose leading orientation dependence is proportional to $\sin2\phioff$ (Figure~\ref{fig:main-symmetry}).

\subsection{Scope and limitations}
\label{subsec:discussion-scope}

The approximation tested here is
\begin{equation}
 \avg{\mathcal{O}(\kappa,\gamma_+, \gamma_\times)}
 \simeq
 \mathcal{O}\!\left(\avg{\kappa},\avg{\gamma_+},0\right)
 \label{eq:discussion-validated-step}
\end{equation}
for an infinitesimally thin, complete, uniformly weighted ring evaluated at one source redshift about a fixed adopted center.  Our numerical tests further impose the conservative subcriticality criterion in Equation~\eqref{eq:main-safety}.

The accuracy of this approximation does not imply that the effects of halo triaxiality or centering offsets on azimuthally averaged profiles or inferred halo properties are small.  Both can change the azimuthally averaged profiles of the lensing fields $\bm F$ and thereby affect fitted masses or concentrations \citep[e.g.,][]{Clowe2004,BeckerKravtsov2011,Grandis2021,Sommer2022,Sommer2024}.  These profile changes enter both sides of Equation~\eqref{eq:discussion-validated-step}: the ring average of the local nonlinear observable and the mean-field prediction constructed from $\avg{\bm F}$.  The residual studied here instead isolates the additional nonlinear difference between them that arises from the angular fluctuations $\delta\bm F$ around the ring (Figure~\ref{fig:main-ring-averaging}).

Within this scope, the results support the use of this mean-field approximation in one-dimensional shear-and-magnification analyses, including its application in the CLUMI and CLUMI+ mass-reconstruction frameworks \citep{Umetsu2013,Umetsu2025}.

A real background sample also spans a distribution of source redshifts.  In this work, each $z_s$ is treated independently, and $\avg{\cdot}$ denotes only an azimuthal average.  Averaging a nonlinear observable over lensing efficiency is a separate noncommuting operation and generally cannot be represented by evaluating the observable at a mean source redshift or mean $\beta$ \citep[e.g.,][]{Umetsu2020rev}.  Quantifying this effect requires the source-redshift distribution and source-dependent weights and is beyond the present scope.

Finite radial bins, angular masks, azimuthally varying source weights, and discrete source sampling are likewise excluded from the ideal-ring calculation.  These effects change the averaging operation and can couple radial or angular structure into the measured estimator.  Their impact thus needs to be evaluated for the specific observational estimator rather than inferred from the complete-ring calculation alone.

Our population experiment contains smooth triaxial host halos with the specified distributions of centering offsets.  It does not include bound subhalos, correlated surrounding structure, or uncorrelated large-scale structure along the line of sight.  If these components are included, the term linear in the angular fluctuations $\delta\bm F$ still vanishes identically on a complete ring, provided the Taylor expansion remains valid.  The present calculation does not determine the angular covariance entries and higher angular moments of that total field, including correlations between the smooth host and the additional structure, or the resulting population scatter.  Quantifying these effects requires a more complete population model.

The adopted offset distributions are intended to characterize the response rather than to calibrate a particular centering proxy.  Predictions for X-ray, SZ, optical, or BCG centers require the corresponding offset distribution, its dependence on halo state, and any alignment with projected shape.  Simulations show that X-ray and SZ offsets can have preferred directions \citep{Sommer2024,Sommer2025}.  The response derived here can be combined with such calibrated offset distributions and alignments.

Finally, the present analysis isolates one general approximation used in one-dimensional weak-lensing inference rather than validating a complete reconstruction procedure.  Method-specific priors, radial modeling, source selection, treatment of the measurement covariance, and parameter degeneracies require separate validation.

\subsection{Relation to previous work}
\label{subsec:discussion-originality}

The monopole relation for tangential shear on a complete ring is well established and does not require axisymmetry \citep{Kaiser1995}.  The corresponding quasi-circular approximation for reduced shear has been used in cluster-lensing analyses \citep[e.g.,][]{Clowe2004,Umetsu2020rev}.  The quadrupole response of elliptical halos has likewise been developed for measuring projected halo shapes with weak lensing \citep{Adhikari2015,ClampittJain2016}.

Using cluster halos from cosmological hydrodynamical simulations, \citet{Grandis2021} directly quantified the difference between averaging the local reduced shear and forming reduced shear from the azimuthally averaged fields.  In a recent simulation-based analysis of cluster weak-lensing mass bias, \citet{Bocquet2026} explicitly computed the two-dimensional reduced-shear field before azimuthal averaging. Separately, \citet{Sommer2024,Sommer2025} showed that randomizing the directions of centering offsets defined from X-ray and SZ observables can change the inferred weak-lensing mass bias, demonstrating that the offset directions need not be statistically independent of projected cluster structure.

The present work addresses the origin and leading structure of this nonlinear ring-averaging residual.  We show how complete-ring cancellation, rotational symmetry, and harmonic orthogonality constrain its dependence on projected shape and miscentering, including their leading orientation-dependent interaction.  
We also study this nonlinear ring-averaging residual for inverse magnification, magnification, magnification bias, and reduced cross shear.

\section{Summary}
\label{sec:summary}

We have asked why the ring average of a nonlinear observable in halo lensing, particularly for cluster-mass halos, can be accurately approximated by the mean-field prediction formed from the ring-averaged lensing fields, even though the field-to-observable mapping does not generally commute with azimuthal averaging.

For the local lensing fields $\bm F(R,\phi)=(\kappa,\gamma_+,\gamma_\times)$, we studied
$\Delta_\phi\mathcal{O}(R)=\avg{\mathcal{O}[\bm F]}-\mathcal{O}[\avg{\bm F}]$,
the difference between averaging the local observable around a ring and evaluating it from the ring-averaged fields.  Both terms are evaluated at the same radius, about the same adopted center, and for the same source plane. Changes in the azimuthally averaged lensing profiles caused by triaxiality or miscentering are already included in both terms, so that $\Delta_\phi\mathcal{O}$ isolates the additional effect of nonlinear averaging (see Figure~\ref{fig:main-ring-averaging}).

Our main conclusions are:
\begin{enumerate}
\item Complete-ring averaging removes the term linear in the angular fluctuations $\delta\bm F=(\delta\kappa,\delta\gamma_+,\delta\gamma_\times)$, whose components scale as $O(\eta\epsilon)$. Where the Taylor expansion is valid, the mean-field prediction is the zeroth-order term and the leading residual is the Hessian--covariance contraction in Equation~\eqref{eq:main-residual-hierarchy}, with higher-order terms beginning at $O(\delta F^3)$. The magnitude and sign of the leading residual therefore depend jointly on the local curvature of the observable and the angular covariance of the lensing fields. For observables with $\mathcal{O}_{\mathrm{MF}}=O(\eta)$, such as reduced shear, the fractional residual scales as $\delta_{\mathcal{O}}=O(\eta\epsilon^2)$; a mean-field baseline close to zero can amplify the fractional residual (Section~\ref{subsec:main-first-order}). 

\item For reduced tangential shear, the $(\kappa,\gamma_+)$ block of the Hessian is indefinite (Equation~\eqref{eq:appA-g-hessian}). Accordingly, the difference between $\avg{g_+(R,\phi)}$ and $g_{+,\mathrm{MF}}(R)$ depends at leading order on the convergence variance and its covariance with tangential shear and can have either sign. For a scalar lensing potential, $\avg{\gamma_\times}=0$ on a complete ring, while $\avg{g_\times}$ need not vanish because reduced shear couples $\gamma_\times$ to angular variations in $\kappa$ (Sections~\ref{subsec:main-reduced-shear} and \ref{subsec:main-cross}).

\item Inverse magnification $Y=\mu^{-1}$ is quadratic, with positive curvature in $\kappa$ and negative curvature in the two shear components, so that $\avg{Y}-\YMf=\Ckk-\Cplusplus-\Ccrosscross$ contains no higher field orders. Magnification and magnification bias apply the additional map $Y\mapsto Y^q$. At second order, Equation~\eqref{eq:main-Yq-second-order} separates the fractional residual for $Y^q$ into a term proportional to $\Delta_\phi Y$ and a variance term whose coefficient $q(q-1)/2$ is set by the curvature of this map. The curvature fixes the sign of the variance term, while the total residual has no universal sign because $\Delta_\phi Y$ can be positive or negative. At all orders, Equation~\eqref{eq:main-Yq-factorization} separates the shift from $\YMf$ to $\avg{Y}$ from the additional effect of applying $Y\mapsto Y^q$ before ring averaging (Section~\ref{subsec:main-magnification}).

\item For a centered elliptical halo, the residual of a scalar observable is even in signed projected ellipticity $\epsp$ when the average covers a complete ring.  Miscentering generates leading $m=1$ structure with local parameter $u=d/R$, whereas projected shape begins at $m=2$ (Figures~\ref{fig:main-ring-averaging} and \ref{fig:main-symmetry}).  Their orthogonality removes the nominal $O(\epsp u)$ term.  The first orientation-dependent terms in $\deltag$ and $\avg{g_\times}$ scale as $O(\epsp u^2\cos2\phioff)$ and $O(\epsp u^2\sin2\phioff)$, respectively (Sections~\ref{subsec:main-evenness}--\ref{subsec:main-interaction}; see Equations~\eqref{eq:main-full-interaction} and \eqref{eq:main-cross-interaction}).

\item For the reference offset model $\sigma_d=0.05$ in the triaxial NFW population experiment (Section~\ref{sec:population} and Table~\ref{tab:main-population-config}), we calculated, for each observable, source plane, population, and radius, the median of the fractional residuals across the retained halos. At $z_s=1$, the largest magnitude of these medians over the six $(\Mtwo,z_l)$ populations and the sampled radii in the common radial domain remains below $1\%$ for each of the five observables (Figure~\ref{fig:main-envelope} and Table~\ref{tab:main-residuals}). Repeating the same realizations at $z_s=2$ increases this largest magnitude for every observable, consistent with weak-lensing power counting, while the corresponding values remain below $2\%$ for all five observables (Section~\ref{subsec:main-source-redshift-test}; see Figure~\ref{fig:main-source-redshift} and Table~\ref{tab:main-residuals}).

\item Across the three nonzero offset models, the hybrid residual predictor in Equation~\eqref{eq:main-closure-predictor} accurately reproduces the reduced-shear residuals $\deltag$ of individual halos for $u=d/R<0.3$, with $\mathcal{R}^2_{\min}=0.9944$ and $\mathcal{R}^2_{\mathrm{med}}=0.9989$ (Section~\ref{subsec:main-large-u}).  At larger $u$, higher-order shape--centering interaction terms become important.  
A $g_{+,\mathrm{MF}}$ value close to zero can separately amplify the fractional residual, while the distinct boundaries $1-\kappa=0$ and $Y=0$ are excluded by the subcriticality criterion (Sections~\ref{subsec:main-conditioning} and \ref{subsec:main-criticality}).  The alignment reweighting recovers the predicted $m=2$ response, and the hybrid predictor remains accurate for the tested aligned populations (Section~\ref{subsec:main-alignment}).
\end{enumerate}

The controlled experiment uses complete, uniformly weighted rings and treats each source plane separately. The far-background subcriticality criterion provides a common mask across source planes and a consistently defined subcritical comparison domain. The halo populations follow prescribed triaxial NFW and centering-offset models, with offset distributions not tied to a particular observational centering proxy.

For the reference offset model $\sigma_d=0.05$ at $z_s=1$, the largest magnitude of the population median for inverse magnification over the six populations and the common radial domain is $0.0093\%$, the smallest of the five observable-specific extrema. This small value reflects population-dependent cancellation between the convergence and shear variance contributions rather than quadraticity alone. For magnification bias with $\alpha=0.3$ in the depletion regime, the corresponding extremal magnitude is $0.069\%$, suggesting that depletion measurements \citep[e.g.,][]{Umetsu2014} are relatively insensitive to this form of nonlinear averaging within the subcritical domain tested here.

Our results provide a physical basis for the mean-field approximation in halo lensing and identify the angular field moments that govern its leading corrections. This offers a systematic route to determining how much angular structure must be modeled to predict nonlinear radial observables to a specified accuracy. Extending the framework to realistic lensing simulations and observational estimators would allow the accuracy of this approximation to be assessed against the precision required for halo mass reconstruction and weak-lensing mass calibration.

\begin{acknowledgements}
This work is supported by the National Science and Technology Council of Taiwan (grants NSTC 112-2112-M-001-027-MY3 and NSTC 115-2112-M-001-027-) and the Academia Sinica Investigator Award (grant AS-IA-112-M04). During preparation of this manuscript, the author used OpenAI's \href{https://openai.com/index/chatgpt/}{ChatGPT} and \href{https://openai.com/index/introducing-codex/}{Codex} as assistive tools for algebraic and numerical cross-checks, code development and review, and editorial revision.

A reference implementation for reproducing the numerical calculations and validation diagnostics presented in this work is publicly available at \href{https://github.com/umetic/halo-lensing-azimuthal-averaging}{https://github.com/umetic/halo-lensing-azimuthal-averaging} (version v0.1.0).
\end{acknowledgements}

\software{NumPy \citep{NumPy2020}, SciPy \citep{SciPy2020}, pandas \citep{McKinney2010}, Matplotlib \citep{Hunter2007}, Colossus \citep{Diemer2018}}

\appendix
\makeatletter
\@addtoreset{equation}{section}
\makeatother
\setcounter{equation}{0}
\renewcommand{\theequation}{\Alph{section}\arabic{equation}}
\renewcommand*{\theHequation}{\Alph{section}.\arabic{equation}}

\section{Additional operator relations and special cases for nonlinear ring averaging}
\label{app:general-observables}

\subsection{Second-order angular covariance entries}
\label{appA:covariance}

Using the covariance notation introduced in Equation~\eqref{eq:main-CAB-definition}, the six independent angular covariance entries of $\bm F=(\kappa,\gamma_+,\gamma_\times)$ on a ring of radius $R$ can be collected into the symmetric matrix
\begin{equation}
 \bm C(R)\equiv
 \begin{pmatrix}
 \Ckk & \Ckplus & \Ckcross\\
 \Ckplus & \Cplusplus & \Cpluscross\\
 \Ckcross & \Cpluscross & \Ccrosscross
 \end{pmatrix}.
 \label{eq:appA-covariance-matrix}
\end{equation}
The mixed angular covariance entries $\Ckcross$ and $\Cpluscross$ are odd under reflection of the local polar basis.  They vanish for a reflection-symmetric lens, while their ensemble expectations vanish in a population for which every configuration and its mirror image have equal statistical weight; neither need vanish for a general individual realization.

\subsection{Reduced-shear Hessian and residual bound}
\label{appA:reduced-shear}

For $g_+=\gamma_+/(1-\kappa)$ and $D=1-\kavg$, the nonzero second derivatives at the mean fields are
\begin{equation}
 g_{+,\kappa\kappa}
 =\frac{2\gpavg}{D^3},
 \qquad
 g_{+,\kappa+}=g_{+,+\kappa}=\frac{1}{D^2}.
 \label{eq:appA-g-hessian}
\end{equation}
Substitution into the Hessian--covariance term in Equation~\eqref{eq:main-residual-hierarchy} gives the expansion of the residual $\Delta_\phi g_+$ in Equation~\eqref{eq:main-delta-g} and the corresponding second-order fractional prediction $\deltag^{(2)}$ in Equation~\eqref{eq:main-deltag-second-order}. Positive semidefiniteness of $\bm C(R)$ implies $|\Ckplus|\leq\sqrt{\Ckk\Cplusplus}$ and therefore
\begin{equation}
 |\Delta_\phi g_+|
 \leq
 \frac{\sqrt{\Ckk\Cplusplus}}{D^2}
 +\frac{|\gpavg|\Ckk}{|D|^3}
 +O(\delta F^3).
 \label{eq:appA-delta-g-bound}
\end{equation}
The bound constrains the magnitude of $\Delta_\phi g_+$ but does not determine its sign. Under Equation~\eqref{eq:main-eta-epsilon}, the $\Ckplus$ contribution to $\Delta_\phi g_+$ is generically $O(\eta^2\epsilon^2)$, whereas the term proportional to $\gpavg\Ckk$ is $O(\eta^3\epsilon^2)$. Since $g_{+,\mathrm{MF}}=O(\eta)$, these two terms contribute $O(\eta\epsilon^2)$ and $O(\eta^2\epsilon^2)$, respectively, to the fractional residual $\deltag$. If the $\Ckplus$ term vanishes, the latter becomes the leading contribution.

\subsection{Reduced-cross variance}
\label{appA:cross}

The azimuthal average of $g_\times$ is given by Equation~\eqref{eq:main-reduced-cross}.  Its leading angular variance induced by lensing is
\begin{equation}
 \mathrm{Var}_\phi(g_\times)
 =\frac{\Ccrosscross}{D^2}+O(\delta F^3).
 \label{eq:appA-gcross-variance}
\end{equation}
The angular variance of $g_\times$ is thus a quadratic diagnostic of non-axisymmetric lensing structure. However, it does not universally predict the leading contribution to $\Delta_\phi g_+$, which depends on $\Ckk$ and $\Ckplus$.

\subsection{Magnification cumulants and the Gaussian case}
\label{appA:magnification}

Writing $Z=\ln Y$, the cumulant identity in Equation~\eqref{eq:main-CGF} may be expanded as
\begin{equation}
 \ln\frac{\avg{Y^q}}{\bigl(\avg{Y}\bigr)^q}
 =\sum_{n=2}^{\infty}
 \frac{q^n-q}{n!}
 \operatorname{cum}_{n,\phi}(\ln Y),
 \label{eq:appA-CGF-series}
\end{equation}
where $\operatorname{cum}_{n,\phi}(Z)\equiv\left.d^nK_Z(t)/dt^n\right|_{t=0}$ denotes the $n$th cumulant of $Z(\phi)$ under the uniform azimuthal average on the ring at fixed $R$ and $z_s$; in particular, $\operatorname{cum}_{1,\phi}(Z)=\avg{Z}$ and $\operatorname{cum}_{2,\phi}(Z)=\mathrm{Var}_\phi(Z)$. The sum begins at $n=2$ because the first-cumulant contribution cancels identically. At second order, $\operatorname{cum}_{2,\phi}(\ln Y)=\mathrm{Var}_\phi(\ln Y)=\mathrm{Var}_\phi(Y)/\YMf^2+O(\delta F^3)$, so that the $n=2$ term reproduces the variance contribution in Equation~\eqref{eq:main-Yq-second-order}. The concavity/convexity statement in Section~\ref{subsec:main-magnification} follows directly from Jensen's inequality and does not assume a particular angular distribution.

If $Z=\ln Y$ is Gaussian, all cumulants above second order vanish and
\begin{equation}
 \frac{\avg{Y^q}}{\bigl(\avg{Y}\bigr)^q}
 =
 \exp\!\left[
 \frac{q(q-1)}{2}\mathrm{Var}_\phi(\ln Y)
 \right].
 \label{eq:appA-lognormal}
\end{equation}
The Gaussian assumption is used only to obtain Equation~\eqref{eq:appA-lognormal}; Equation~\eqref{eq:main-CGF} holds for any angular distribution of $Z=\ln Y$.

\section{Potential multipoles, coherent quadrupole, and elliptical-profile closure}
\label{app:coherent-halo}

\subsection{Real potential modes and consistent field responses}
\label{appB:potential}

The angular structure can be formulated at the level of the standard dimensionless scalar lensing potential \citep{SchneiderBartelmann1997,BernsteinNakajima2009}.  In this subsection and the potential-level closure below we use the angular radius $\theta$; the physical projected radius used for cluster profiles is $R=D_l\theta$, as defined in Section~\ref{subsec:main-setup}.  Write
\begin{equation}
 \psi(\theta,\phi)=\psimono(\theta)
 +\sum_{m=1}^{\infty}
 [a_m(\theta)\cos m\phi+b_m(\theta)\sin m\phi].
 \label{eq:appB-potential-expansion}
\end{equation}
Here $\psimono$ is the azimuthally symmetric monopole of the potential.  No restriction is imposed on the radial dependence or phase of a mode.  For each $m$, define the radial operators
\begin{equation}
 \mathcal{D}^{\kappa}_m \equiv
 \frac12\left(\frac{d^2}{d\theta^2}+\frac1\theta\frac{d}{d\theta}-\frac{m^2}{\theta^2}\right),
 \quad \mathcal{D}^{+}_m \equiv
 -\frac12\left(\frac{d^2}{d\theta^2}-\frac1\theta\frac{d}{d\theta}+\frac{m^2}{\theta^2}\right),
 \quad \mathcal{D}^{\times}_m \equiv
 m\left(\frac1\theta\frac{d}{d\theta}-\frac1{\theta^2}\right).
 \label{eq:appB-response-operators}
\end{equation}
With $\bm\Psi_m=(a_m,b_m)^T$ and the quarter-turn matrix
\begin{equation}
 \bm J=\begin{pmatrix}0&-1\\1&0\end{pmatrix},
\end{equation}
the cosine--sine coefficient vectors of the fields are
\begin{equation}
 \bm F_m^\kappa=\mathcal{D}^{\kappa}_m\bm\Psi_m,
 \qquad
 \bm F_m^+=\mathcal{D}^{+}_m\bm\Psi_m,
 \qquad
 \bm F_m^\times=\mathcal{D}^{\times}_m\bm J\bm\Psi_m.
 \label{eq:appB-field-coefficients}
\end{equation}
Thus convergence, tangential shear, and cross shear are not independent phenomenological expansions; they are linked differential responses to the same scalar-potential modes.

For the centered halo in the principal-axis frame of Section~\ref{subsec:main-shape}, the coefficient vectors are $\bm F_2^\kappa=(F_2^\kappa,0)^T$, $\bm F_2^+=(F_2^+,0)^T$, and $\bm F_2^\times=(0,F_2^\times)^T$, with the signed scalar amplitudes defined in Equation~\eqref{eq:main-m2-amplitudes}.

We define the contribution of mode $m$ to the two-radius angular covariance kernel by
\begin{equation}
 P_m^{AB}(\theta,\theta')
 \equiv
 \frac12\bm F_m^A(\theta)\cdot\bm F_m^B(\theta'),
 \qquad A,B\in\{\kappa,+,\times\}.
 \label{eq:appB-mode-kernel}
\end{equation}
Angular orthogonality gives
\begin{equation}
 \avg{\delta F_A(\theta,\phi)\delta F_B(\theta',\phi)}
 =\sum_{m=1}^{\infty}P_m^{AB}(\theta,\theta').
 \label{eq:appB-kernel-sum}
\end{equation}
At equal radii, $\theta'=\theta$, Equation~\eqref{eq:appB-kernel-sum} gives the angular covariance entry $C_{AB}(R)$ after the coordinate relabeling $R=D_l\theta$.  The same formalism can in principle organize coherent and stochastic angular components by physical origin, although the population experiment models only the coherent effects of shape and centering in smooth halos.  An additive power budget would require ensemble averaging together with explicit independence assumptions.

\subsection{Ring average for a coherent quadrupole}
\label{appB:quadrupole}

Following Section~\ref{subsec:main-shape}, we measure $\phi$ from the projected major axis.  We consider a centered halo for which the only angular variation is the $m=2$ mode:
\begin{equation}
 \kappa =\kavg+F_2^\kappa\cos2\phi, 
  \quad \gamma_+ = \gpavg+F_2^+\cos2\phi,
  \quad \gamma_\times = F_2^\times\sin2\phi.
 \label{eq:appB-quadrupole-fields}
\end{equation}
The angular covariance entries are
\begin{equation}
 \Ckk=\frac12(F_2^\kappa)^2,
 \quad
 \Ckplus=\frac12F_2^+F_2^\kappa,
 \quad
 \Cplusplus=\frac12(F_2^+)^2,
 \quad
 \Ccrosscross=\frac12(F_2^\times)^2,
 \quad
 \Ckcross=\Cpluscross=0.
 \label{eq:appB-quadrupole-operators}
\end{equation}
When $D>|F_2^\kappa|$, the reduced-shear denominator is positive at every azimuth, and the exact ring average of the reduced tangential shear is
\begin{equation}
 \avg{g_+}
 =
 \frac{\gpavg}{S_2}
 +\frac{F_2^+F_2^\kappa}{S_2(D+S_2)},
\quad
S_2\equiv\sqrt{D^2-(F_2^\kappa)^2}.
 \label{eq:appB-exact-g}
\end{equation}
The corresponding residual is
\begin{equation}
 \Delta_\phi g_+
 =
 \frac{F_2^\kappa}{S_2(D+S_2)}
 \left(F_2^++\frac{\gpavg F_2^\kappa}{D}\right).
 \label{eq:appB-exact-delta-g}
\end{equation}
Expanding for $|F_2^\kappa|/D<1$ gives
\begin{equation}
 \Delta_\phi g_+
 =
 \frac{F_2^+F_2^\kappa}{2D^2}
 +\frac{\gpavg (F_2^\kappa)^2}{2D^3}
 +O(\epsp^4),
 \label{eq:appB-quadrupole-expansion}
\end{equation}
which reproduces the general second-order operator result.  For $\gpavg\neq0$, dividing the second-order terms by $g_{+,\mathrm{MF}}=\gpavg/D$ gives the centered-shape contribution
\begin{equation}
 \deltag^{\mathrm{ell},(2)}(R\mid\epsp)
 =
 \frac{F_2^+F_2^\kappa}{2D\gpavg}
 +\frac{(F_2^\kappa)^2}{2D^2}.
 \label{eq:appB-deltag-ell-second-order}
\end{equation}
The leading $O(\epsp)$ forms of $F_2^\kappa$ and $F_2^+$ for the area-preserving elliptical profile are given in Appendix~\ref{appB:elliptical-profile}.

For this centered quadrupole, reflection symmetry makes $\avg{g_\times}$ vanish identically.  In the formal limit $D\rightarrow|F_2^\kappa|^+$, $S_2\rightarrow0$.  This zero-denominator limit is excluded by the subcriticality criterion; see Section~\ref{subsec:main-criticality} for the convergence discussion.

\subsection{Nonlinear mode coupling in inverse magnification}
\label{appB:m4}

Substituting the coherent-quadrupole fields in Equation~\eqref{eq:appB-quadrupole-fields} into the inverse magnification defined in Equation~\eqref{eq:main-Y-definitions} gives
\begin{equation}
 Y(R,\phi)
 =\avg{Y}+Y_2\cos2\phi+Y_4\cos4\phi,
 \label{eq:appB-Y-modes}
\end{equation}
with
\begin{equation}
 \begin{aligned}
 \Delta_\phi Y
 &\equiv\avg{Y}-\YMf
 =\frac12\left[(F_2^\kappa)^2-(F_2^+)^2-(F_2^\times)^2\right],\\
 Y_2&=-2(DF_2^\kappa+\gpavg F_2^+),\\
 Y_4&=\frac12\left[(F_2^\kappa)^2-(F_2^+)^2+(F_2^\times)^2\right].
 \end{aligned}
 \label{eq:appB-Y-coefficients}
\end{equation}
Thus, a pure $m=2$ perturbation in the lensing fields shifts the ring mean by $\Delta_\phi Y$ and generates an $m=4$ component in the nonlinear inverse-magnification field.  Orthogonality gives
\begin{equation}
 \mathrm{Var}_\phi(Y)
 =\frac12Y_2^2+\frac12Y_4^2
 =2(DF_2^\kappa+\gpavg F_2^+)^2+O(\epsp^4),
 \label{eq:appB-Y-variance}
\end{equation}
and
\begin{equation}
 \mathrm{Var}_\phi(\ln Y)
 =\frac{2(DF_2^\kappa+\gpavg F_2^+)^2}{\YMf^2}
 +O(\epsp^4).
\end{equation}
Using $\Delta_\phi Y$, the ring-averaged $Y^q$ response through second order is
\begin{equation}
 \begin{aligned}
 \frac{\avg{Y^q}}{\YMf^q}
 &=
 1+q\frac{\Delta_\phi Y}{\YMf}
 +\frac{q(q-1)}{\YMf^2}
  (DF_2^\kappa+\gpavg F_2^+)^2
 +O(\epsp^4),\\
 \left.
 \frac{\avg{Y^q}}{\YMf^q}
 \right|_{q=1}
 &=
 1+\frac{\Delta_\phi Y}{\YMf}
 =1+\deltaY.
 \end{aligned}
 \label{eq:appB-quadrupole-Yq}
\end{equation}
At $q=1$, the map $Y\mapsto Y^q$ is linear, so that the variance term and all higher-order corrections generated by this map vanish.
The second line of Equation~\eqref{eq:appB-quadrupole-Yq} thus recovers the inverse-magnification residual in Equation~\eqref{eq:main-deltaY-exact}.

\subsection{Closure by a single potential amplitude}
\label{appB:potential-closure}

If the quadrupole is generated by
\begin{equation}
 \psi(\theta,\phi)=\psimono(\theta)+a_2(\theta)\cos2\phi,
\end{equation}
the field amplitudes are
\begin{equation}
 F_2^\kappa = \frac12\left(a_2''+\frac{a_2'}\theta-\frac{4a_2}{\theta^2}\right),
 \quad F_2^+ = -\frac12\left(a_2''-\frac{a_2'}\theta+\frac{4a_2}{\theta^2}\right),
 \quad F_2^\times = 2\left(\frac{a_2'}\theta-\frac{a_2}{\theta^2}\right),
 \label{eq:appB-potential-amplitudes}
\end{equation}
where primes in this subsection denote derivatives with respect to $\theta$.  They obey the compatibility relation
\begin{equation}
 F_2^\kappa-F_2^+=F_2^\times+\frac{\theta}{2}\frac{dF_2^\times}{d\theta}.
 \label{eq:appB-compatibility}
\end{equation}
The three amplitudes are thus linked but not generally equal.

\subsection{Area-preserving elliptical surface density}
\label{appB:elliptical-profile}

We now return to the physical projected radius $R=D_l\theta$, which is the natural coordinate for cluster surface-density profiles.  Since $D_l$ is constant for a fixed lens, logarithmic radial derivatives are unchanged by the conversion between $R$ and $\theta$, and the dimensionless field amplitudes above can be relabeled as functions of $R$.  The leading multipoles below are equivalent to the standard quadrupole formulas for elliptical halos \citep{Adhikari2015,ClampittJain2016,Umetsu2020rev}.  With $e=(1-q_\perp^2)/(1+q_\perp^2)$, one has $e/2=\epsp+O(\epsp^3)$; integration by parts converts the usual radial integrals to the form used here.  For the circular projected profile $\ksph(R)$ of the centered spherical reference halo, deformed into self-similar area-preserving elliptical contours, define
\begin{equation}
 q_\perp=\frac{1-\epsp}{1+\epsp},
 \qquad
 \zeta^2=q_\perp x^2+\frac{y^2}{q_\perp},
 \qquad
 \kappa(R,\phi)=\ksph(\zeta).
 \label{eq:appB-elliptical-radius}
\end{equation}
The small-ellipticity expansion is
\begin{equation}
 \frac{\zeta}{R}
 =1-\epsp\cos2\phi
 +\epsp^2\left(\frac34-\frac14\cos4\phi\right)
 +O(\epsp^3).
 \label{eq:appB-zeta-expansion}
\end{equation}
Extending the definition of $F_2^\kappa$ in Equation~\eqref{eq:main-m2-amplitudes}, we define the $m=4$ convergence coefficient by
\begin{equation}
 F_4^\kappa(R)\equiv2\avg{\delta\kappa(R,\phi)\cos4\phi}.
 \label{eq:appB-K4-definition}
\end{equation}
Writing $\kappa_{\mathrm{sph},1}\equiv d\ksph/d\ln R$ and $\kappa_{\mathrm{sph},2}\equiv d^2\ksph/d(\ln R)^2$, the leading convergence multipoles are
\begin{equation}
 \begin{aligned}
 \kavg&=\ksph+\frac{\epsp^2}{4}(\kappa_{\mathrm{sph},2}+2\kappa_{\mathrm{sph},1})+O(\epsp^4),\\
 F_2^\kappa&=-\epsp\kappa_{\mathrm{sph},1}+O(\epsp^3),\\
 F_4^\kappa&=\frac{\epsp^2}{4}(\kappa_{\mathrm{sph},2}-2\kappa_{\mathrm{sph},1})+O(\epsp^4).
 \end{aligned}
 \label{eq:appB-density-multipoles}
\end{equation}
We also define the fourth-order interior radial moment of the spherical convergence profile by
\begin{equation}
 \overline{\kappa}_4(R)
 \equiv
 \frac4{R^4}\int_0^R dR'\,R'^3\ksph(R').
 \label{eq:appB-kappa4}
\end{equation}
The leading $m=2$ shear amplitudes are
\begin{equation}
 F_2^+ = \epsp[\kappa_{\mathrm{sph},1}-2\ksph+3\overline{\kappa}_4]+O(\epsp^3),
 \quad F_2^\times = \epsp[-4\ksph+3\overline{\kappa}_4]+O(\epsp^3).
 \label{eq:appB-shear-responses}
\end{equation}
The convergence quadrupole is local in the density slope, whereas the shear quadrupoles retain nonlocal information about the interior mass distribution.  Equations~\eqref{eq:appB-density-multipoles} and \eqref{eq:appB-shear-responses} thus reproduce the standard tangential- and cross-shear quadrupoles in our ellipticity convention.  Here they are used to evaluate nonlinear ring averages.  \citet{Fu2026} retain second-order terms in ellipticity when modeling the monopole and quadrupole of stacked cluster lensing.

For a power-law projected profile $\ksph(R)\propto R^{-n}$, the leading amplitudes are
\begin{equation}
 F_2^\kappa=\epsp n\ksph+O(\epsp^3),
 \quad F_2^+=\epsp\frac{n^2-2n+4}{4-n}\ksph+O(\epsp^3),
 \quad F_2^\times=4\epsp\frac{n-1}{4-n}\ksph+O(\epsp^3).
 \label{eq:appB-powerlaw-amplitudes}
\end{equation}
For $0<n<2$, the leading reduced-shear coefficient is positive.  In the weak-lensing limit,
\begin{equation}
 \deltag
 \simeq
 \epsp^2
 \frac{(2-n)(n^2-2n+4)}{2(4-n)}\ksph
 +O(\epsp^4\ksph)
 +O(\epsp^2\ksph^2).
 \label{eq:appB-powerlaw-deltag}
\end{equation}
This expression makes explicit the double suppression by projected ellipticity and lensing strength.

Axis interchange satisfies
\begin{equation}
 \zeta(R,\phi;-\epsp)
 =\zeta\left(R,\phi+\frac\pi2;\epsp\right),
\end{equation}
which proves the even-power theorem stated in Equation~\eqref{eq:main-even-ellipticity}.

For a numerical consistency test, we use a dimensionless projected NFW profile with $\kappa_s=0.12$ and $q_\perp=0.67$ ($\epsp\simeq0.20$), adopting the same elliptical NFW construction as in Figure~\ref{fig:main-symmetry}.  Here $\kappa_s\equiv\rho_s r_s/\Sigma_{\mathrm{cr}}$, where $\rho_s$ and $r_s$ are the NFW characteristic density and scale radius; with this convention, $\ksph(r_s)=2\kappa_s/3$.  The normalization $\kappa_s=0.12$ is chosen solely to provide a safely subcritical configuration and does not match that of the physical single-halo model in that figure; the entire domain $0.2\leq R/r_s\leq5$ remains safely subcritical.  

At $R=r_s$, direct evaluation from the local lensing fields gives the reduced-shear residual $\deltag=2.442\times10^{-3}$, compared with $2.400\times10^{-3}$ from Equation~\eqref{eq:appB-deltag-ell-second-order}, with $D$ and $\gpavg$ evaluated for the adopted elliptical profile.  The analogous second-order predictions for $\deltaY$ and the magnification-bias residuals with $\alpha=0.3$ and $1.4$ agree with the directly evaluated residuals to within $5.6\%$ of the residual magnitude.  The remaining differences are consistent with the expected $O(\epsp^4)$ truncation, and higher even modes carry only a few percent of the nonmonopole angular power in this moderately subcritical example.

\section{Triaxial projection, population model, and numerical methods}
\label{app:population-numerics}

\subsection{Projection of a volume-preserving triaxial NFW halo}
\label{appC:triaxial}

A density profile stratified on similar triaxial ellipsoids projects to a surface-density profile stratified on similar ellipses \citep{Stark1977,Binney1985}.  The corresponding triaxial NFW lensing formalism has been developed and applied extensively \citep[e.g.,][]{Oguri2005,SerenoUmetsu2011,Umetsu2015}.  Here we write these standard projection relations in a compact matrix form adapted to our volume-preserving parameterization.

Let $p\equiv b/a$ and $s\equiv c/a$ denote the intrinsic intermediate-to-major and minor-to-major axis ratios, respectively, with $0<s\leq p\leq1$.  We construct the triaxial halo by applying a volume-preserving affine deformation to a spherical reference profile.  The stretch factors along the intrinsic major, intermediate, and minor axes are
\begin{equation}
 A=(ps)^{-1/3},
 \qquad
 B=pA,
 \qquad
 C=sA,
 \qquad
 ABC=1.
\end{equation}
An isodensity surface of ellipsoidal radius $m$ thus has semiaxes $(Am,Bm,Cm)$.  Since $ABC=1$, it encloses the same volume as a sphere of radius $m$.

Let $\bm x=(x_1,x_2,x_3)^T$ denote the three-dimensional position relative to the halo center, expressed in the intrinsic principal-axis frame, and let $\rho_{\mathrm{sph}}(r)$ be the spherical reference NFW density profile.  The density of the triaxial halo is then
\begin{equation}
 \rho_{\mathrm{tri}}(\bm x)=\rho_{\mathrm{sph}}(m),
 \qquad
 m^2=\bm x^T\bm S\bm x,
 \qquad
 \bm S=\operatorname{diag}(A^{-2},B^{-2},C^{-2}).
\end{equation}
Thus, the mass enclosed by the ellipsoid of radius $m$ equals the mass enclosed within radius $m$ in the spherical reference halo.

In the same principal-axis frame, let $\hat{\bm n}$ be the unit vector along the line of sight, and let the columns of the $3\times2$ matrix $\bm E$ form an orthonormal basis for the sky plane perpendicular to $\hat{\bm n}$.  Writing $\bm x=\bm E\bm X+t\hat{\bm n}$, where $\bm X$ is the two-dimensional sky-plane coordinate in this basis and $t$ is the line-of-sight coordinate, and completing the square in $t$ gives
\begin{equation}
 m^2
 =\xi_{\mathrm{los}}\left(t+\frac{\bm v^T\bm X}{\xi_{\mathrm{los}}}\right)^2
 +\bm X^T\bm P\bm X,
\end{equation}
where
\begin{equation}
 \xi_{\mathrm{los}}=\hat{\bm n}^{T}\bm S\hat{\bm n},
 \qquad
 \bm v=\bm E^T\bm S\hat{\bm n},
 \qquad
 \bm P=\bm E^T\bm S\bm E-\frac{\bm v\bm v^T}{\xi_{\mathrm{los}}}.
\end{equation}
Here $\xi_{\mathrm{los}}$ determines the line-of-sight scaling, while the symmetric positive-definite matrix $\bm P$ defines the projected elliptical contours.

Let $\Sigma_{\mathrm{sph}}(R)$ denote the projected surface-density profile of the spherical reference halo.  Integration along the line of sight gives
\begin{equation}
 \Sigma_{\mathrm{tri}}(\bm X)
 =b_{\mathrm{los}}\,\Sigma_{\mathrm{sph}}
 \!\left(\sqrt{\bm X^T\bm P\bm X}\right),
 \qquad
 b_{\mathrm{los}}=\xi_{\mathrm{los}}^{-1/2}.
 \label{eq:appC-exact-projection}
\end{equation}
If $\lambda_1\leq\lambda_2$ are the eigenvalues of $\bm P$, the projected axis ratio is $q_\perp=\sqrt{\lambda_1/\lambda_2}$.  Since $\det\bm S=(ABC)^{-2}=1$, the Schur-complement identity gives $\det\bm P=\xi_{\mathrm{los}}^{-1}=b_{\mathrm{los}}^2$.  Hence $\lambda_1=b_{\mathrm{los}}q_\perp$ and $\lambda_2=b_{\mathrm{los}}/q_\perp$.  In the projected principal-axis frame, let $X_{\mathrm{maj}}$ and $X_{\mathrm{min}}$ denote coordinates along the projected major and minor axes and define the elliptical radius
\begin{equation}
 \zeta^2=q_\perp X_{\mathrm{maj}}^2+\frac{X_{\mathrm{min}}^2}{q_\perp}.
\end{equation}
Since the two coefficients have unit product, a contour of constant $\zeta$ encloses an area $\pi\zeta^2$, independent of $q_\perp$.  Thus, $\zeta$ defines an area-preserving elliptical radius.  We then have $\bm X^T\bm P\bm X=b_{\mathrm{los}}\zeta^2$.

Defining the circular projected reference profile by
\begin{equation}
 \Sigma_{\mathrm{ref}}(R)
 \equiv b_{\mathrm{los}}\,\Sigma_{\mathrm{sph}}\!\left(\sqrt{b_{\mathrm{los}}}R\right),
\end{equation}
we can write the projected triaxial surface-density field as
\begin{equation}
 \Sigma_{\mathrm{tri}}(X_{\mathrm{maj}},X_{\mathrm{min}})=\Sigma_{\mathrm{ref}}(\zeta).
\end{equation}
Thus, each projected realization is an area-preserving elliptical deformation of a circular NFW reference profile, in the same form used in Appendix~\ref{appB:elliptical-profile}.  The corresponding projected NFW parameters are
\begin{equation}
 r_{s,\perp}=b_{\mathrm{los}}^{-1/2}r_s,
 \qquad
 \kappa_{s,\perp}=b_{\mathrm{los}}\kappa_s.
 \label{eq:appC-projected-parameters}
\end{equation}
Here $\kappa_s$ and $\kappa_{s,\perp}$ denote the spherical and projected NFW convergence normalizations for a fixed source plane.  For a given projected realization, setting $q_\perp=1$ at fixed circular reference profile therefore removes the projected ellipticity without changing its radial scale or lensing normalization, as used in the offset-only construction in Equation~\eqref{eq:main-shape-offset-exact}.  Direct numerical integration along the line of sight verifies Equation~\eqref{eq:appC-exact-projection} with fractional differences below $10^{-12}$ for random intrinsic shapes and viewing orientations.

\subsection{Mass, concentration, and shape distributions}
\label{appC:population}

We parameterize each triaxial NFW halo by its spherical-equivalent mass and concentration,
$\Mtwo=(4\pi/3)200\rho_{\mathrm{c}}(z)\rtwo^3$ and $c_{200\mathrm c}=\rtwo/r_s$,
where $\rtwo$ is the three-dimensional spherical overdensity radius and $r_s$ is the NFW scale radius.  We denote projected lensing radii by $R$.  The volume-preserving deformation changes the halo shape while leaving $\Mtwo$ and $c_{200\mathrm c}$ unchanged.

At fixed mass and redshift, our baseline population samples halo concentration, intrinsic axis ratios, and viewing direction independently.  Concentrations are lognormally distributed about a median mass--concentration relation, intrinsic shapes follow the fitting prescription of \citet{Bonamigo2015} evaluated at cluster-scale peak heights, and viewing directions are isotropic.  The model does not include correlations of halo shape with formation history or selection-dependent weighting of viewing orientation \citep[e.g.,][]{Lau2021}.

Specifically, we draw $\log_{10}[c_{200\mathrm c}/c^{\mathrm{DJ19}}_{200\mathrm c,\,\mathrm{med}}(\Mtwo,z)]\sim\mathcal{N}(0,0.16^2)$, corresponding to a lognormal scatter of $0.16$~dex about the median relation of \citet{DiemerJoyce2019}, which we evaluate with \textsc{Colossus} \citep{Diemer2018}.

As defined in Appendix~\ref{appC:triaxial}, the intrinsic intermediate-to-major and minor-to-major axis ratios are $p=b/a$ and $s=c/a$, respectively, with $0<s\leq p\leq1$.  Their mass and redshift dependence enters through the linear-theory peak height
$\nu\equiv\delta_{\mathrm{c}}/\sigma(M_{\mathrm{vir}},z)$, where $\delta_{\mathrm{c}}$ is the linear collapse threshold and $\sigma(M_{\mathrm{vir}},z)$ is the rms linear density fluctuation on the virial mass scale.  We evaluate $\nu$ with \textsc{Colossus} at the virial mass corresponding to each $(\Mtwo,z)$ grid point, using the median concentration to convert between mass definitions.

We define $\widetilde s\equiv s\nu^{0.255}$ and draw
$\ln\widetilde s\sim\mathcal{N}(-0.49,0.20^2)$, rejecting draws that give $s\geq1$.  
Conditional on $s$, we define $\widetilde p\equiv(p-s)/(1-s)$ and draw
$\widetilde p\sim\operatorname{Beta}(\alpha_{\widetilde p},\beta_{\widetilde p})$, with
\begin{equation}
 \mu_{\widetilde p}=0.633s-0.007,
 \quad
 \beta_{\widetilde p}=1.389s^{-1.685},
 \quad
 \alpha_{\widetilde p}
 =\beta_{\widetilde p}
 \frac{\mu_{\widetilde p}}{1-\mu_{\widetilde p}}.
 \label{eq:appC-bonamigo-p}
\end{equation}
The physical intermediate-to-major axis ratio is then
$p=s+(1-s)\widetilde p$.

Table~\ref{tab:main-population-config} summarizes the numerical configuration.

\subsection{Centering-offset distribution}
\label{appC:offsets}

Define the dimensionless centering offset as $x_d\equiv d/\rtwo$. 
The adopted offset-amplitude distribution is
\begin{equation}
 p(x_d\mid\sigma_d)
 =\frac{x_d}{\sigma_d^2}
 \exp\!\left(-\frac{x_d^2}{2\sigma_d^2}\right),
 \qquad x_d\geq0,
 \label{eq:appC-Rayleigh}
\end{equation}
with an independent offset direction drawn uniformly from $[0,2\pi)$.  Thus $d$ (or $x_d$) varies from halo to halo, whereas $\sigma_d$ specifies the population distribution.  Population summaries include the halo-to-halo distribution of $d$ and are labeled by $\sigma_d$.  For each halo, we draw one unit-Rayleigh variate and one offset direction and reuse them for all nonzero values of $\sigma_d$.  The offset models are thus compared realization by realization.  In the baseline model, the offset amplitude and direction are independent of mass, redshift, concentration, intrinsic shape, and dynamical state.  Appendix~\ref{appD:alignment} introduces alignment between offset direction and projected shape through importance reweighting.

\subsection{Numerical accuracy checks}
\label{appC:validation}

The displaced elliptical NFW fields are evaluated with a Fourier--Green mode expansion over $0.18\leq q_\perp\leq1$ and $10^{-6}\leq R/r_{s,\perp}\leq31$, retaining modes through $m_{\max}=20$ and using $N_\phi=256$ angular samples.  All halo realizations considered here lie within this numerical domain, so that no extrapolation is required.

We verified the reduced-shear calculation against an independent evaluation of displaced halos; the largest absolute difference in $\deltag$ is $7.4\times10^{-9}$.  Independent calculations of the magnification observables at $z_s=1$ and 2 agree to numerical precision.

\section{Miscentering, shape--centering coupling, and alignment}
\label{app:miscentering-alignment}

\subsection{Fields of a displaced spherical halo}
\label{appD:exact-offset}

Consider a spherical halo whose true center is displaced by $d$ along the $x$ direction from the adopted analysis center.  For this spherical reference halo, rotational symmetry allows the offset direction to define the $x$ axis, so that the amplitude $d$ fully specifies the displacement.  We thus write $\bm F(R,\phi\mid d)$ for the local lensing fields evaluated about the adopted center and conditional on that displacement.

On a ring of radius $R$ about the adopted center, the true halo-centric projected radius $R_{\mathrm{h}}$ is given by Equation~\eqref{eq:main-displaced-radius}.  Let $\phi_{\mathrm{h}}$ be the polar angle of the same point about the true halo center, measured from the same $x$ axis, and define $\omega\equiv\phi_{\mathrm{h}}-\phi$.  Thus, $\omega$ is the signed rotation between the two local radial directions.  It satisfies
\begin{equation}
 \cos\omega=\frac{R-d\cos\phi}{R_{\mathrm{h}}},
 \qquad
 \sin\omega=\frac{d\sin\phi}{R_{\mathrm{h}}}.
 \label{eq:appD-rotation-angle}
\end{equation}
For the centered spherical reference profiles $\ksph(R_{\mathrm{h}})$ and $\gsph(R_{\mathrm{h}})$, the fields evaluated about the adopted center are
\begin{equation}
 \kappa(R,\phi\mid d) = \ksph(R_{\mathrm{h}}),
 \quad \gamma_+(R,\phi\mid d) = \gsph(R_{\mathrm{h}})\cos2\omega,
 \quad \gamma_\times(R,\phi\mid d) = \gsph(R_{\mathrm{h}})\sin2\omega.
 \label{eq:appD-exact-offset-fields}
\end{equation}
The derivation below depends only on the circular projected profiles $\ksph$ and $\gsph$, and hence also applies to the fixed circular reference profile associated with each projected triaxial realization.

Reflection symmetry about the offset axis gives
\begin{equation}
 \avg{\gamma_\times(R,\phi\mid d)}
 =\avg{g_\times(R,\phi\mid d)}
 =\Ckcross(R\mid d)=\Cpluscross(R\mid d)=0
\end{equation}
for the displaced spherical halo.

The mean-field baseline $g_{+,\mathrm{MF}}(R\mid d)$ is evaluated about the adopted center according to Equation~\eqref{eq:main-gmf-definition}.  It thus already includes the suppression or reshaping of the radial mean fields relative to the true center.  The corresponding fractional residual is
\begin{equation}
 \deltag^{\mathrm{off}}(R\mid d)
 \equiv
 \frac{\avg{g_+(R,\phi\mid d)}}{g_{+,\mathrm{MF}}(R\mid d)}-1.
 \label{eq:appD-deltag-off}
\end{equation}
Using the first-order small-offset fields in Equation~\eqref{eq:main-m1-fields}, with $u=d/R$ and primes denoting derivatives with respect to $R$, angular averaging gives the leading $O(u^2)$ contributions to the angular covariance entries:
\begin{equation}
\begin{aligned}
 \Ckk^{\mathrm{off}}(R\mid d)&=\frac{d^2}{2}\ksph'^2,
 &\Ckplus^{\mathrm{off}}(R\mid d)&=\frac{d^2}{2}\ksph'\gsph',\\
 \Cplusplus^{\mathrm{off}}(R\mid d)&=\frac{d^2}{2}\gsph'^2,
 &\Ccrosscross^{\mathrm{off}}(R\mid d)&=2\frac{d^2}{R^2}\gsph^2.
\end{aligned}
 \label{eq:appD-offset-operators}
\end{equation}
The omitted terms begin at $O(u^4)$.  Substitution into Equation~\eqref{eq:main-deltag-second-order} gives the second-order offset contribution
\begin{equation}
 \deltag^{\mathrm{off},(2)}(R\mid d)
 =A_u(R)u^2
 =\frac{d^2}{2}\left[
 \frac{\ksph'\gsph'}{(1-\ksph)\gsph}
 +\frac{\ksph'^2}{(1-\ksph)^2}
 \right].
 \label{eq:appD-deltag-off-second-order}
\end{equation}
Provided the fractional residual is analytic in the components of $\bm d$ at $\bm d=0$, rotational invariance permits only even powers of $d$.  At fixed $R$, the residual therefore satisfies $\deltag^{\mathrm{off}}(R\mid d)=\deltag^{\mathrm{off},(2)}(R\mid d)+O(u^4)$.

For $0<n<2$ and $\ksph(R)\propto R^{-n}$, the second-order contribution reduces to
\begin{equation}
 \deltag^{\mathrm{off},(2)}(R\mid d)
 =\frac{n^2}{2}\left(\frac{d}{R}\right)^2
 \frac{\ksph}{(1-\ksph)^2}.
 \label{eq:appD-deltag-off-powerlaw}
\end{equation}
This expression makes the leading quadratic dependence on the local expansion variable $u=d/R$ explicit.

As a numerical benchmark, we evaluate a spherical NFW halo with $M_{200\mathrm c}=10^{15}h^{-1}M_\odot$, $z_l=0.3$, $z_s=1$, and $c_{200\mathrm c}=3.84$.  For this monotonically decreasing profile, the inner edge of the contiguous outer branch satisfying $\lambda_{-,\infty}\geq0.1$ obeys $R_{\mathrm{sc}}(d)=d+R_{\mathrm{sc}}(0)$, with $R_{\mathrm{sc}}(0)=0.0538\rtwo$.  At $R=0.3\rtwo$, direct numerical evaluation of Equation~\eqref{eq:appD-deltag-off} from the local lensing fields gives $\deltag^{\mathrm{off}}=0.147\%$, $0.638\%$, and $3.78\%$ for $d/\rtwo=0.05$, $0.10$, and $0.20$, respectively.  The leading $O(u^2)$ prediction in Equation~\eqref{eq:appD-deltag-off-second-order} has relative errors of approximately $2.7\%$, $10.6\%$, and $39.6\%$ for the same offsets.  At $R=R_{\mathrm{sc}}(d)$, the leading prediction recovers about $74\%$ and $54\%$ of $\deltag^{\mathrm{off}}$ for $d/\rtwo=0.10$ and $0.20$, respectively.  This benchmark confirms the quadratic translation scaling at modest $d/R$ and its progressive breakdown once the translated field becomes strongly nonperturbative.

\subsection{Shape--centering interaction}
\label{appD:interaction}

Let the projected major axis define $\phi=0$ and let the offset direction make angle $\phioff$ with that axis.  Once shape is present, the offset amplitude alone no longer specifies the geometry. We thus write the explicit realization dependence as $(\epsp,d,\phioff)$; the mass, concentration, source redshift, and other fixed quantities remain implicit.
Writing the leading shape amplitudes as $F_2^A=\epsp\widehat F_2^A+O(\epsp^3)$, with $A\in\{\kappa,+,\times\}$, the combined fields to first order in shape and translation are
\begin{equation}
 \begin{aligned}
 \delta\kappa(R,\phi\mid\epsp,d,\phioff)&=\epsp\widehat F_2^\kappa(R)\cos2\phi-d\,\ksph'(R)\cos(\phi-\phioff)+\cdots,\\
 \delta\gamma_+(R,\phi\mid\epsp,d,\phioff)&=\epsp\widehat F_2^+(R)\cos2\phi-d\,\gsph'(R)\cos(\phi-\phioff)+\cdots,\\
 \gamma_\times(R,\phi\mid\epsp,d,\phioff)&=\epsp\widehat F_2^\times(R)\sin2\phi+2u\gsph(R)\sin(\phi-\phioff)+\cdots.
 \end{aligned}
 \label{eq:appD-combined-fields}
\end{equation}

In the population calculations, we evaluate the shape-only and offset-only residuals defined in Equation~\eqref{eq:main-shape-offset-exact} directly from the corresponding local lensing fields for each projected halo.
For $\delta_{g,i}^{\mathrm{ell}}$ we set $d_i=0$.  For $\delta_{g,i}^{\mathrm{off}}$ we set $q_{\perp,i}=1$ while keeping the circular reference profile and offset vector unchanged; the projected radial scale and lensing normalization inherited from the triaxial projection therefore remain fixed.  The displaced elliptical configuration supplies $\delta_{g,i}$.

Complete-ring orthogonality of $m=1$ and $m=2$ removes every $O(\epsp u)$ contribution to the quadratic operators, giving the leading additivity in Equation~\eqref{eq:main-leading-additivity}.

The allowed higher-order terms follow from three transformations:
\begin{equation}
 \begin{aligned}
 \phioff &\rightarrow\phioff+\pi
 \quad\text{(offset reversal)},\\
 \phioff &\rightarrow-\phioff
 \quad\text{(reflection)},\\
 (\epsp,\phioff) &\rightarrow(-\epsp,\phioff+\pi/2)
 \quad\text{(axis interchange)}.
 \end{aligned}
\end{equation}
A parity-even scalar thus has the leading expansion in Equation~\eqref{eq:main-full-interaction}, whereas $\avg{g_\times}$, which is parity odd, has the form in Equation~\eqref{eq:main-cross-interaction}.  Averaging over a uniform offset direction removes the $O(\epsp u^2\cos2\phioff)$ term; the orientation-averaged interaction begins at $O(\epsp^2u^2)$.

For the illustrative configuration with $q_\perp=0.67$ and $d=0.1\rtwo$ at $R=0.3\rtwo$, direct evaluation of $\deltag$ from the displaced elliptical lensing fields gives $1.21\%$, $0.980\%$, and $0.742\%$ for offsets along the major axis, at $45^\circ$, and along the minor axis, respectively. The corresponding shape-only and offset-only residuals are $0.304\%$ and $0.638\%$, respectively.
The orientation interaction thus enhances or suppresses the leading additive response with the predicted sign.

At small offsets, the scaling is particularly clean: at $R=0.3\rtwo$, the fitted $\cos2\phioff$ and $\sin2\phioff$ amplitudes divided by $d^2$ vary by only a few percent over $d/\rtwo=0.005$--$0.04$.  Translation of an elliptical halo also induces higher harmonics, but $m=1$ and $m=2$ remain dominant while the low-order expansion is accurate.

\subsection{Offset--major-axis alignment}
\label{appD:alignment}

We implement the offset--major-axis alignment tests of Section~\ref{subsec:main-alignment} by importance reweighting the halo realizations.  We use the far-background lensing fields throughout these tests.  The calculation grid comprises the six $(\Mtwo,z_l)$ grid points, the three nonzero centering-offset models $\sigma_d=0.02$, $0.05$, and $0.08$, and the four scaled radii $R/\rtwo$ in Equation~\eqref{eq:main-closure-radii}.

At each configuration, we keep the halo realizations and offset amplitudes fixed and change the distribution of $\phioff$ through importance reweighting.  The baseline parent distribution is uniform, $p_0(\phioff)=1/(2\pi)$.  An aligned population with target angular density $p(\phioff)$ is obtained by assigning realization $i$ the importance weight $w_i=p(\varphi_{\mathrm{off},i})/p_0(\varphi_{\mathrm{off},i})$.

Following Equation~\eqref{eq:main-Qoff}, we characterize the target parent distribution by its $m=2$ moment, $\Qoff\equiv\langle\cos2\phioff\rangle$.  Positive $\Qoff$ favors offsets along the projected major axis, while negative $\Qoff$ favors the minor axis.

Our primary angular model is
\begin{equation}
 p_{\mathrm{lin}}(\phioff\mid\Qoff)
 =\frac{1+2\Qoff\cos2\phioff}{2\pi},
 \qquad |\Qoff|\leq\frac12.
 \label{eq:appD-linear-alignment}
\end{equation}
We evaluate $\Qoff=-0.30,-0.15,0,0.15,$ and $0.30$.  

To test sensitivity to higher Fourier moments, we also use the nematic von Mises angular model
\begin{equation}
 p_{\mathrm{VM}}(\phioff\mid\kappa_\varphi)
 =\frac{\exp(\kappa_\varphi\cos2\phioff)}
        {2\pi I_0(\kappa_\varphi)},
 \qquad
 \frac{I_1(\kappa_\varphi)}{I_0(\kappa_\varphi)}=\Qoff,
 \label{eq:appD-vonmises-alignment}
\end{equation}
where $I_n$ is the modified Bessel function of the first kind. For nonzero $\Qoff$, the two angular models have the same target $\Qoff$ but different higher Fourier moments; at $\Qoff=0$, both reduce to the uniform distribution.

Let $S_i^{\mathrm{outer}}(R_j)=1$ if realization $i$ satisfies Equation~\eqref{eq:main-safety} at $R_j$ and at every larger sampled radius, and let $S_i^{\mathrm{outer}}(R_j)=0$ otherwise.  The represented fraction of the reweighted population is
\begin{equation}
 f_{\mathrm{rep},w}^{\mathrm{outer}}(R_j)
 \equiv
 \frac{\sum_i w_i S_i^{\mathrm{outer}}(R_j)}
 {\sum_i w_i}.
 \label{eq:appD-weighted-represented-fraction}
\end{equation}
We use this weighted represented fraction when applying the 95\% representation requirement to an aligned population.  Finite Monte Carlo sampling and any dependence of the far-background subcriticality selection on offset direction can cause the empirical $m=2$ moment of the retained weighted sample to differ from the target $\Qoff$.

For each $(\Mtwo,z_l,\sigma_d,R/\rtwo)$ configuration and selected $u$ domain, we first determine $a_0$ and $a_{\mathrm{int}}$ in Equation~\eqref{eq:main-closure-predictor} from the baseline uniform-angle population, whose parent distribution has $\Qoff=0$.  We then apply these coefficients to the same halo realizations reweighted to nonzero $\Qoff$.  This construction tests whether a predictor fitted to random offset directions remains accurate for aligned populations.

Using the reduced-shear residuals $y_i$ defined in Equation~\eqref{eq:main-closure-target} and the predictions $\widehat y_i$ evaluated with these fixed coefficients, we define the weighted coefficient of determination by
\begin{equation}
 \mathcal{R}_w^2
 \equiv
 1-\frac{\sum_i w_i(y_i-\widehat y_i)^2}
         {\sum_i w_i(y_i-y_{\mathrm{mean},w})^2},
 \qquad
 y_{\mathrm{mean},w}
 \equiv
 \frac{\sum_i w_i y_i}{\sum_i w_i}.
 \label{eq:appD-weighted-R2}
\end{equation}
For this predictor test, the sums run over the retained halos in the selected $u$ domain.  An overall normalization of the weights cancels, and equal weights recover Equation~\eqref{eq:main-R2-definition}.  Weighted means and quantiles refer to the empirical distribution defined by the same $w_i$.

We quantify the statistical support of the importance reweighting using the effective sample size
\begin{equation}
 N_{\mathrm{eff}}
 \equiv
 \frac{(\sum_i w_i)^2}{\sum_i w_i^2}.
 \label{eq:appD-effective-sample-size}
\end{equation}
For a selected sample containing $N$ halos before reweighting, equal weights give $N_{\mathrm{eff}}=N$.  Across all domains satisfying the 95\% representation requirement in Equation~\eqref{eq:appD-weighted-represented-fraction}, $N_{\mathrm{eff}}/N \gtrsim 0.84$ even for the strongest alignment considered here.

Table~\ref{tab:main-alignment-response} reports the weighted-median alignment response at $R=0.292\rtwo$ for the reference offset model, $\sigma_d=0.05$.  The offset-scale, angular-model, and predictor comparisons are reported in Section~\ref{subsec:main-alignment}.  Both angular models are reflection symmetric, giving $\langle\sin2\phioff\rangle=0$.  Consistent with Equation~\eqref{eq:main-cross-interaction}, the weighted population mean of the ring-averaged reduced cross shear remains consistent with zero, although its scatter among halos changes.

\FloatBarrier
\bibliographystyle{aasjournalv7}

\end{document}